\documentclass[twocolumn,prl,superscriptaddress,showpacs]{revtex4-1} 
\usepackage{amsmath,amssymb}
\usepackage{graphicx}
\usepackage{verbatim}
\usepackage{amsfonts}
\usepackage{subfigure}
\usepackage{dcolumn}
\usepackage{float}
\usepackage{bm}
\usepackage{color}
\usepackage[colorlinks,citecolor=blue]{hyperref}
\usepackage{framed} 
\usepackage{mathrsfs}
\usepackage{ntheorem}
\usepackage{tikz-cd}
\usepackage{longtable}
\usepackage{booktabs}
\usepackage{hyperref}
\usepackage{natbib}
\usepackage{empheq}
\usepackage{multirow}
\usepackage{enumitem}
\usepackage[normalem]{ulem} 
\useunder{\uline}{\ul}{}
\usetikzlibrary{cd}

\newcommand{\beq}{\begin{eqnarray} }
	\newcommand{\eeq}{\end{eqnarray} }
\newcommand{\Beq}{\begin{eqnarray*} }
	\newcommand{\Eeq}{\end{eqnarray*} }
\newcommand{\Bmat}{\left(\begin{matrix}}
	\newcommand{\Emat}{\end{matrix}\right)}

\newcommand{\bit}{\begin{itemize} }
\newcommand{\eit}{\end{itemize} }
\newcommand{\ben}{\begin{enumerate} }
\newcommand{\een}{\end{enumerate} }

\definecolor{lightblue}{rgb}{0, 0.439, 0.753}

\newcommand{\mi}{\mathrm{i}}

\begin{document}

\title{On the Symmetries of Anisotropic Spin Interaction Models}

\author{Arist Zhenyuan Yang}
\email{zhenyuan@hku.hk}
\affiliation{Department of Physics and HK Institute of Quantum Science \& Technology, The University of Hong Kong, Pokfulam Road, Hong Kong, China}

\date{\today}

\begin{abstract}

We show that anisotropic spin interactions do not merely break spin-space group (SSG) symmetries, but instead \emph{twist} them through cohomology invariants, yielding symmetry classes beyond  subgroups of \( O(3)\times \operatorname{Isom}(\mathbb{R}^3) \). This requires redefining the spin-only group \( \mathcal{S}_0 \) in terms of proper spin rotations. Based on this unitary \( \mathcal{S}_0 \), we formulate a twisted SSG (tSSG) theory that captures the complete set of spin-space symmetries. We then study a spin-1 model with tSSG symmetry using linear flavor wave theory and find $\mathbb{Z}_2$ topological quadrupolar excitations defined on a spin Brillouin Klein bottle. Specifically, the quadrupolar excitations possess a momentum-space glide-reflection symmetry and the edge states exhibit a nonlocal momentum twist. These results establish the symmetry language required for interacting spin systems, whether realized in magnetic materials or on programmable quantum simulators, and open a route to unconventional magnetism.

\end{abstract}

\maketitle

\paragraph*{Introduction.—} Spin-space groups (SSGs) provide a unifying symmetry framework for magnetic materials, enabling systematic descriptions of magnetic topological electronic states~\cite{Liu2022SpinGroupPRX,Yang2024SymmetryInvariantsNatCommun,Guo2021EightfoldPRL,Ghorashi2024AltermagneticMajoranaPRL}, topological magnon bands~\cite{KatsuraNagaosaLee2010ThermalHall,chumak2015magnon,Chisnell2015TopologicalMagnonKagome,cheng2016spin,Li2017DiracNodalMagnonsPRL,Yao2018TopologicalSpinExcitationsNatPhys,CorticelliMoessnerMcClarty2022MagnonBandTopology,CorticelliMoessnerMcClarty2023ComplexMagnonTopology,chen2025unconventional}, and the altermagnetism~\cite{Smejkal2022EmergingPRX,Cheong2022MagneticChiralityNPJQM,Bose2022TiltedSpinCurrentNatElectron,Feng2022AHEAltermagneticRuO2NatElectron,Karube2022SpinSplitterTorquePRL,Bai2022SpinSplittingTorquePRL,Smejkal2022BeyondConventionalPRX,Mazin2022EditorialAltermagnetismPRX}. These systems are governed by isotropic Heisenberg exchange in which spin and spatial operations combine in a particularly transparent manner: each symmetry operation is realized as a global spin rotation in $O(3)$ accompanied by a lattice isometry, and the resulting SSGs form subgroups of $O(3)\times \operatorname{Isom}(\mathbb{R}^3)$~\cite{BrinkmanElliottPeierls1966SpinSpaceGroups,LitvinOpechowski1974SpinGroups,jiang2024enumeration,chen2024enumeration,Xiao2024SpinSpaceGroupsPRX,song2025constructions}.

In realistic magnets, anisotropic spin interactions are ubiquitous and often essential~\cite{Halilov1998MAEFeCoNi,Nagaosa2013SkyrmionReview,Fert2013SkyrmionsOnTheTrack,nembach2015linear,Winter2017GeneralizedKitaevMagnetism,Takagi2019KitaevQSLReview}.
Canonical examples include Dzyaloshinskii-Moriya (DM) exchange~\cite{Dzyaloshinsky1958WeakFerromagnetism,Moriya1960AnisotropicSuperexchange}, compass-type bond interactions~\cite{DorierBeccaMila2005QuantumCompass,Kitaev2006Anyons,JackeliKhaliullin2009HeisenbergToKitaev,NussinovVanDenBrink2015CompassModels}, and single-ion anisotropy originating from the crystal field~\cite{sachidanandam1997single,craig20153d,meng2016understanding}. Moreover, these terms are natural targets of the quantum simulation of magnetism: programmable platforms based on trapped ions, optical lattices, and Rydberg arrays have realized a variety of interacting spin models~\cite{FriedenauerSchmitzGlueckertPorrasSchaetz2008QuantumMagnet,SimonBakrMaTaiPreissGreiner2011AntiferromagneticSpinChains,BlochDalibardNascimbene2012UltracoldQuantumSimulations,GeorgescuAshhabNori2014QuantumSimulation,BrowaeysLahaye2020RydbergManyBody,MonroeEtAl2021TrappedIonSpinSimulations}, while polar molecules provide a direct route to long-range, tunable anisotropic exchange~\cite{MicheliBrennenZoller2006PolarMoleculesSpinModels,GorshkovEtAl2011UltracoldPolarMoleculesQuantumMagnetism,YanMosesGadwayCoveyHazzardReyJinYe2013DipolarSpinExchange}.

Spin anisotropies are commonly assumed to reduce or preserve the symmetry of isotropic magnets~\cite{schiff2025crystallographic,LiuChenYuEtAl2026OrientedSpinSpaceGroups}. However, we propose the model [Eq.~(\ref{The-model})] to show that anisotropy instead changes the \emph{type} of symmetry realized: exact symmetries of anisotropic models involve spatially varying spin rotations, so that the full symmetry group cannot embed into \( O(3)\times \mathrm{Isom}(\mathbb{R}^3) \). We further find that these spin symmetries acquire a precise group-theoretic meaning within group extension theory only after removing effective time-reversal from the spin-only group \( \mathcal{S}_0 \).

Motivated by these findings, we formulate a theory of twisted spin-space groups (tSSGs) that completes the symmetry description of interacting spin systems via group extension theory. The key step is to redefine \( \mathcal{S}_0 \) as the group of on-site \emph{proper} spin rotations that are exact symmetries of the system. With this definition, we show that any spin-space operation $g$ admits a natural decomposition into (i) a lattice operation $l_g$ (unitary or antiunitary), (ii) a global normalizer spin operation $R_g\in O(3)$ that implements the global conjugation action on $\mathcal{S}_0$, and (iii) a site-dependent centralizer sector $\mathfrak{Z}_g(\vec{r})$ that commutes with $\mathcal{S}_0$ at every site. If \( \mathfrak{Z}_g(\vec r)\equiv 1\), then $g\in O(3)\times \operatorname{Isom}(\mathbb{R}^3)$, and such operations necessarily form a subgroup of $O(3)\times \operatorname{Isom}(\mathbb{R}^3)$. In contrast, when there exists an element $g$ with nontrivial  centralizer sector $\mathfrak{Z}_g(\vec r)$, the group structure, as will be shown below,  is modified by a two cocycle \( [\omega_2] \in H^2_{\phi}(G_{\mathcal L}, \mathcal Z(\mathcal S_0)) \). Here \( G_{\mathcal L} \) denotes the set of lattice operations \( \{l_g,\forall g\} \) and \( \phi: G_{\mathcal L} \to \operatorname{Aut}(\mathcal Z(\mathcal S_0)) \) is the action induced by conjugation on the center of \( \mathcal{S}_0 \) via the normalizer sector $R_g$.

Remarkably, tSSGs give rise to a novel class of topological magnetic excitations on nonorientable manifolds. We study a two-dimensional anisotropic spin-1 model [Eq.~(\ref{quadrupolarH})] on the wallpaper group \( pm \) using linear flavor wave theory. The Hamiltonian contains uniform XYZ anisotropy and DM interactions on \( x \) bonds, uniform off-diagonal symmetric anisotropy and DM interactions on \( y \) bonds, and single-ion anisotropy \( D(S_i^z)^2 \).  In the large-\(D\) regime, the quadrupolar sector \(\{Q_{xz},Q_{yz}\}\) possesses a momentum-space glide-reflection symmetry; that is, the corresponding Hamiltonian obeys the sewing relation
\[
U H_{\mathrm{BdG}}(k_x,k_y) U^{-1} = H_{\mathrm{BdG}}(-k_x,k_y+\pi).
\]
Consequently, the corresponding topological excitations are characterized not on the conventional Brillouin torus, but on a Brillouin Klein bottle~\cite{chen2022brillouin,zhang2023general,ZhangYangZhao2025ProjectiveCrystalSymmetry,zhang2025brillouin}. In particular, the two edge modes are related by the nonlocal twist $\omega_L(k_y)=\omega_R(k_y+\pi)$. These features sharply distinguish them from conventional topological magnetic excitations.

We also present a general scheme for models realizing tSSGs based on the cocycle free construction of group extensions, and provide a complete classification of spin point groups with \( \mathcal{S}_0 \) valued in the 11 chiral point groups, summarized in the Supplementary Material (SM).

\paragraph*{A minimal model with twisted collinear SSG.—} The model is defined on a two-dimensional square lattice with spatial symmetry $p4mm$. It contains nearest-neighbor $XY$ interactions along the $\vec{x}$ direction and $DM$ interactions along the $\vec{y}$ direction. Specifically, for an undirected bond $\langle ij\rangle \parallel \vec{x}$, $H^{XY}_{\langle ij\rangle}= S_i^x S_j^x + S_i^y S_j^y$, whereas for a directed bond $\langle k\!\to\! l\rangle \parallel \vec{y}$, $ H^{DM}_{\langle k\to l\rangle}= S_k^x S_l^y - S_k^y S_l^x $. Both interactions have equal strength $\alpha$, a fixed nonzero real constant. The coupling to the AFM background field is described by $H_a=\sum_l (-1)^l h S_l^z $ with $h$ its strength. The total Hamiltonian then reads

\begin{small}
\begin{align}\label{The-model}
H\!\!=\!\!\sum_{\langle ij\rangle}\!\left(\alpha\, H_{\langle ij\rangle}^{XY}\!+\!S^z_{i}S^z_{j}\right)\!+\!\!\sum_{\langle k\to l\rangle}\!\!\left( \alpha\, H^{DM}_{\langle k\to l\rangle} \!+\! S^z_{k}S^z_{l}\right)+H_{a}.
\end{align}
\end{small}

We first analyze the unitary symmetries of the model. Since the space group (SG) $p4mm$ is symmorphic, it admits the decomposition $p4mm \equiv \mathcal{C}_{4v}\ltimes(\mathbb{T}_x\times\mathbb{T}_y)$, where $\mathcal{C}_{4v}$ is generated by $\{C_4,\mathcal{M}_x\}$. The Wyckoff position $1b=(1/2,1/2)$ is fixed by $\mathcal{C}_{4v}$, and $\mathbb{T}_x$ and $\mathbb{T}_y$ denote translations along the $x$ and $y$ directions, respectively. With this notation, the Hamiltonian is invariant under five global unitary operations: spin rotations $SO(2)_z=\{R_z(\theta)\}$, two effective translations $(R_x(\pi)\mid\mid\mathbb{T}_{x,y})$, and two effective mirror reflections $(R_x(\pi)\mid\mid\mathcal{M}_{x,y})$. These symmetries are all contained in collinear SSGs.

\begin{figure}[b]
  \centering
\includegraphics[scale=0.21]{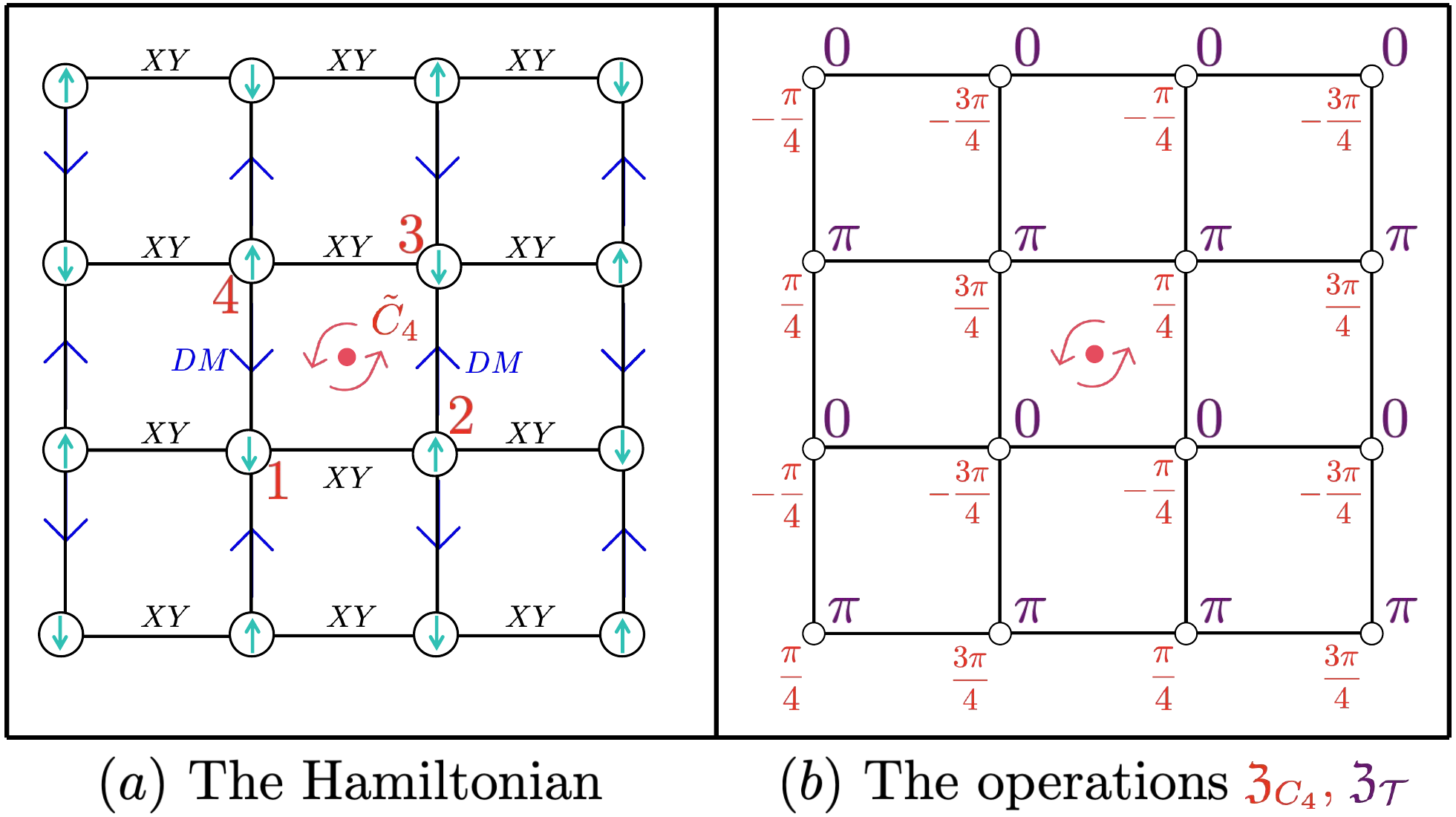}
\caption{Spin model with twisted collinear SSG symmetries $\widetilde{C}_4$ and $\widetilde{\mathcal{T}}$. Panel (a) illustrates the Hamiltonian and the AFM background field
(teal), while panel (b) shows the site-dependent spin operations associated with ${C}_4$ (red) and ${\mathcal{T}}$ (purple), respectively. Here, each $\theta\in[0,2\pi)$ in the panel (b) stands for a local spin rotation $R_{z}(\theta)\in SO(2)_z$.}\label{model-tSSG}
\end{figure}

An exotic unitary symmetry arises when the lattice rotation $C_4$ is spin-decorated to become a symmetry of the Hamiltonian. The resulting effective rotation necessarily exchanges the $XY$ interaction on $x$-bonds with the $DM$
interaction on $y$-bonds,
\begin{align}
\widetilde{C}_{4}:\quad x\text{-bond } XY \;\longleftrightarrow\; y\text{-bond } DM .
\end{align}
This transformation cannot be realized by lattice $C_4$ followed by any global spin rotation. Nevertheless, owing to the local gauge equivalence between the $XY$ and $DM$ terms~\cite{shekhtman1992moriya,yildirim1995anisotropic}, it can be implemented by introducing a site-dependent spin operation $\mathfrak{Z}_{C_4}(\vec r)$, see Fig.~\ref{model-tSSG}. On a single square, it reads

\begin{small}
\begin{align}
&\mathfrak{Z}_{C_{4}}({ r}_{1})\!\cdot\!\vec{S}_{{ r}_1}\!=\!\vec S_{{ r}_1} R_{z}\left(-{3\pi\over 4}\right),\;\mathfrak{Z}_{C_{4}}({ r}_{2})\!\cdot\!\vec S_{{ r}_2}\!=\!\vec S_{{r}_2} R_{z}\left(-{\pi\over 4}\right)\notag\\
&\mathfrak{Z}_{C_{4}}({ r}_{3})\!\cdot\!\vec{S}_{{ r}_3}\!=\!\vec S_{{ r}_3}R_{z}\left({\pi\over 4}\right),\;\;\;\;\;\;\,\mathfrak{Z}_{C_{4}}({ r}_{4})\!\cdot\!\vec S_{{ r}_4}\!=\!\vec S_{{r}_4} R_{z}\left({3\pi\over 4}\right). \notag
\end{align}\end{small}The combined operation $\widetilde{C}_{4}\equiv(\mathfrak{Z}_{C_{4}}(\vec{r}) R_{x}(\pi)\mid\mid C_4)$ which satisfies
\begin{align}\label{GroupStructuretildeC4}
\widetilde{C}_4SO(2)_z(\widetilde{C}_4)^{-1}=(SO(2)_z)^{-1},\;\; (\widetilde{C}_4)^4=R_{z}(\pi),
\end{align}
is a symmetry of the Hamiltonian (\ref{The-model}).

We now demonstrate the nontrivial nature of $\widetilde{C}_4$ in comparison with conventional spin point groups (SPGs). The key result is as follows: there exist \emph{exactly three} distinct groups $\mathcal{G}$ satisfying the quotient relation $\mathcal{G}/SO(2)_z \cong \mathcal{C}_4 $~\cite{note1}. Among these, only two correspond to conventional SPGs, while the third is intrinsically distinct and is realized by the operation $\widetilde{C}_4$ obeying Eq.~(\ref{GroupStructuretildeC4}).

The two conventional SPGs correspond to the direct product $\mathcal{G}_1 \equiv SO(2)_z \times \mathcal{C}_4$ and the semidirect product
$\mathcal{G}_2 \equiv SO(2)_z \rtimes \mathscr{C}_4$, where $\mathscr{C}_4$ is generated by $\{(R_x(\pi)\mid\mid C_4)\}$. To see this, we recall several basic facts about collinear SSGs~\cite{jiang2024enumeration,chen2024enumeration,Xiao2024SpinSpaceGroupsPRX}. From the classification theory, a collinear SSG $G_{cl}$ is characterized by the decomposition $G_{cl} = G'_{cl} \times \big(SO(2)_z \rtimes \mathbb{Z}_2^{\mathcal{T}'}\big)$, where $\mathbb{Z}_2^{\mathcal{T}'}$ is generated by ${\mathcal{T}'}=( -R_x(\pi)\mid\mid \mathcal{T}  )$ with $\mathcal{T} $ the time-reversal. Here, $O(2)_z \equiv SO(2)_z \rtimes \mathbb{Z}_2^{\mathcal{T}'}$ is referred to as the spin-only group. Each element $g' \in G'_{cl}$ takes the form $(E\mid\mid b_{g'})$ or $( -E\mid\mid b_{g'}\mathcal{T})$, with $b_{g'}$ a spatial operation. Let $G_{\mathcal{L}}$ denote the SG generated by all such $b_{g'}$. Collinear SSGs are then in one-to-one correspondence with one-dimensional real representations of $G_{\mathcal{L}}$.

For our purposes, we assume that the lattice symmetry group $G_{\mathcal{L}}$ contains the cyclic subgroup $\mathcal{C}_4$. Restricting $G_{\mathcal{L}}$ to $\mathcal{C}_4$, there are exactly two SPG solutions, corresponding to the two one-dimensional real representations of $\mathcal{C}_4$. In the first case, $G'_{cl}$ is generated by the unitary operation $(E\mid\mid C_4)$; in the second, it is generated by the antiunitary operation $( -E\mid\mid C_4\mathcal{T})$. These two SPG solutions can be rewritten as $ G_{cl}^{(+)} = SO(2)_z \rtimes (\mathcal{C}_4 \times \mathbb{Z}_2)$, with $\mathcal{C}_4$ generated by $(E\mid\mid C_4)$, and $ G_{cl}^{(-)} = SO(2)_z \rtimes (\mathscr{C}_4 \times \mathbb{Z}_2)$, with $\mathscr{C}_4$ generated by the unitary operation $(R_x(\pi)\mid\mid C_4)$. Restricting $G_{cl}^{(+)}$ and $G_{cl}^{(-)}$ to their unitary subgroups yields the direct product group $\mathcal G_1$ and the semidirect product group $\mathcal G_2$, respectively.

We now turn to the group $\mathcal{G}_3$ generated by $SO(2)_z$ and $\widetilde{C}_4$, which satisfies Eq.~(\ref{GroupStructuretildeC4}). Owing to the presence of the spin flipping $R_x(\pi)$ in $\widetilde{C}_4$, $\mathcal{G}_3$ appears superficially similar to $\mathcal{G}_2$. Indeed, the semidirect
product $\mathcal{G}_2$ can be expressed as
\begin{align}\label{GroupStructurescrC4}
\mathscr{C}_4\,SO(2)_z\,\mathscr{C}_4^{-1}=(SO(2)_z)^{-1},\; (\mathscr{C}_4)^4=R_z(0)=E,
\end{align}
where we denote $\{R_x(\pi)\mid\mid C_4\}$ by $\mathscr{C}_4$. The crucial distinction between Eqs.~(\ref{GroupStructurescrC4}) and (\ref{GroupStructuretildeC4}) lies in the fourth power of the corresponding effective $C_4$ operation. Importantly, this distinction cannot be removed: $(\widetilde{C}_4)^4=R_z(\pi)$ is in fact a cohomology invariant. In other words, $\mathcal{G}_2$ cannot be transformed into $\mathcal{G}_3$ by redefining $\mathscr{C}_4$ within $\mathcal{G}_2$. Mathematically, such a redefinition corresponds to a different choice of section and hence induces a coboundary transformation, which leaves the invariant unchanged. We further note that $\mathcal{G}_2$ and $\mathcal{G}_3$ are not only distinct in this mathematical sense, but are also physically inequivalent, as they are not conjugate to each other.

We next analyze the antiunitary symmetries of the model. The effective time-reversal symmetry $\mathcal{T}'=(-R_{x}(\pi)\mid\mid\mathcal{T})$, present in all collinear SSGs, is explicitly broken. Nevertheless, it can be restored by combining it with a site-dependent spin operation $\mathfrak{Z}_{\mathcal{T}}(\vec r)$. On the square in Fig.~\ref{model-tSSG}, we define $ \mathfrak{Z}_{\mathcal{T}}(\vec r_{1,2})=1$, $ \mathfrak{Z}_{\mathcal{T}}(\vec r_{3,4})=R_z(\pi)$, under which the Hamiltonian is invariant with respect to $ \widetilde{\mathcal{T}}=\big(-\mathfrak{Z}_{\mathcal{T}}(\vec r)\,R_x(\pi)\mid\mid\mathcal{T}\big)$.

The site-dependent spin operations $\mathfrak{Z}_{g}(\vec r)$ in $\widetilde{C}_4$ and $\widetilde{\mathcal{T}}$ modify the cohomology invariants and generate symmetry groups absent in the conventional formulation. This mechanism originates from the nontrivial conjugation of $SO(2)_z$ by $R_x(\pi)$, which necessitates decomposing the full symmetry group with respect to the unitary subgroup $SO(2)_z$, rather than the spin-only group $O(2)_z$. A decomposition based on $O(2)_z$ instead obscures this conjugation structure and the associated cohomology invariants. This choice is further enforced by the fact that the conventional antiunitary operation $\mathcal{T}'\in O(2)_z$ is broken and replaced by the site-dependent symmetry $\widetilde{\mathcal{T}}$.

\paragraph*{Basic Setup.—} Below, we redefine the spin-only group and formulate a basic conceptual framework for symmetries in spin systems. Detailed constructions of tSSGs are presented in the following section.

A spin-only group is defined as a subgroup of \(SO(3)\) and is denoted by \(\mathcal{S}_0\). Representative examples of \(\mathcal{S}_0\) include the 11 chiral point groups (cPGs) consisting solely of proper rotations: 6 abelian groups \(\mathcal{C}_1\), \(\mathcal{C}_2\), \(\mathcal{C}_3\), \(\mathcal{C}_4\), \(\mathcal{C}_6\), and \(\mathcal{D}_2\), and 5 nonabelian groups \(\mathcal{D}_3\), \(\mathcal{D}_4\), \(\mathcal{D}_6\), \(\mathcal{T}\), and \(\mathcal{O}\). Continuous groups such as \(SO(2)_z,SO(3)\) also fall into this category. In the following, we denote by \(\mathcal{Z}(\mathcal{S}_0)\) the center of \(\mathcal{S}_0\), and by \(\mathfrak{N}_{\mathcal{S}}(\mathcal{S}_0)\) and \(\mathfrak{Z}_{\mathcal{S}}(\mathcal{S}_0)\) its normalizer and centralizer in a group \(\mathcal{S}\subset O(3)\), respectively.

For a given spin Hamiltonian, the spin-only group $\mathcal{S}_0$ is defined as the set of all on-site global proper rotations that leave the Hamiltonian invariant. By definition, $\mathcal{S}_0$ is a normal subgroup of the full symmetry group $\mathcal{G}$. Then $\mathcal{G}$ admits a coset decomposition,
\begin{align}\label{Gcoset}
\mathcal{G}= \mathcal{S}_0 \sqcup u_1 \mathcal{S}_0 \sqcup u_2 \mathcal{S}_0 \sqcup \cdots \sqcup T_1 \mathcal{S}_0 \sqcup T_2 \mathcal{S}_0 \sqcup \cdots ,
\end{align}
where $u_1, u_2, \ldots$ are unitary elements and $T_1, T_2, \ldots$ are antiunitary elements. These cosets themselves form a group isomorphic to a magnetic space group (MSG) $G_{\mathcal{L}}$, equivalently $\mathcal{G}/\mathcal{S}_0 \cong G_{\mathcal{L}}$. Conversely, the complete class $\mathscr{G}$ of symmetry groups for spin systems with a given $\mathcal{S}_0$ is naturally defined as
\begin{align}\label{DefinetSSG}
\mathscr{G} = \left\{ \mathcal{G} \,\middle|\, \mathcal{G}/\mathcal{S}_0 \cong \mathrm{MSG} \right\}.
\end{align}
The set $\mathscr{G}$ splits into two classes. One class consists of groups that can be embedded as subgroups of the direct product $O(3)\times\operatorname{Isom}(\mathbb{R}^3)$; the other does not admit such an embedding. All elements of $\mathscr{G}$ are referred to as tSSGs, including cases with nontrivial cohomology twists such as $\mathcal{G}_3$ defined in Eq.~(\ref{GroupStructuretildeC4}). Groups belonging to the first class are, by convention, also referred to as SSGs. Two tSSGs $\mathcal{G}, \mathcal{G}' \in \mathscr{G}$ are defined to be physically equivalent if there exists a spin transformation $s\in O(3)$ such that $\mathcal{G} = s\,\mathcal{G}'\,s^{-1}$. Equivalently, the two groups are related by a global change of spin coordinates.

However, the definition in Eq.~(\ref{DefinetSSG}) is overly general and may include groups $\mathcal{G}$ that are not physically meaningful for spin systems. For example, when $\mathcal{S}_0 = SO(2)_z$, the nontrivial conjugation action on $\mathcal{S}_0$ is not unique to the spin flipping $R_x(\pi)$, but can also be realized by other unitary matrices, such as $U=\operatorname{diag}(\mathrm{i}\sigma_x,1)\in SU(3)$. Allowing such conjugations would lead to an $SU(3)$ theory, which lies outside the scope of  spin physics. We therefore impose the following restriction: \emph{the conjugation actions of any operations in $\mathcal{G}\in\mathscr{G}$ on $\mathcal{S}_0$ must be realized by elements of $O(3)$.} Under this condition, $\mathscr{G}$ defines a complete class of tSSGs relevant to spin systems. In the following, $\mathscr{G}$ refers exclusively to this restricted class.

\paragraph*{Twisted Spin-Space Groups.—} We now construct tSSGs. The construction is abstract, and the representative example introduced above will be used to clarify the key ideas when appropriate. In what follows, $\mathcal{G}$ denotes a tSSG with spin-only group $\mathcal{S}_0$, satisfying $\mathcal{G}/\mathcal{S}_0 \cong G_{\mathcal{L}}$, where $G_{\mathcal{L}}$ is a MSG.

We show that the structure underlying the operation $\widetilde{C}_4=(\mathfrak{Z}_{C_4}R_x(\pi)\mid\mid C_4)$ is generic to all tSSGs. Owing to the global nature of $\mathcal{S}_0$, each element $g\in\mathcal{G}$ admits a systematic decomposition into three sectors. Let $s_g(\vec r_i)\in O(3)$ denote the spin operation implementing the conjugation action of $g$ on $\mathcal{S}_0$ at site $\vec r_i$. Since elements of $\mathcal{S}_0$ act globally, the conjugation action of $g$ preserves $\mathcal{S}_0$ at all sites, implying
\begin{align}
s_g(\vec r_i)\,\mathcal{S}_0\,s_g(\vec r_i)^{-1} \equiv s_g(\vec r_j)\,\mathcal{S}_0\,s_g(\vec r_j)^{-1} = \mathcal{S}_0 .
\end{align}
Thus, $s_g(\vec r_i)$ and $s_g(\vec r_j)$ normalize $\mathcal{S}_0$, and their difference $s_g(\vec r_i)s_g(\vec r_j)^{-1}$ lies in the centralizer of $\mathcal{S}_0$. Fixing a reference site $\vec r_0$, each element $g\in\mathcal{G}$ can therefore be written as
\[
g=\big(\mathfrak{Z}_g(\vec r)\,R_g^0\mid\mid l_g\big),
\]
where $l_g\in G_{\mathcal L}$ is the lattice operation of $g$, $R_g^0$ is the global operation valued in the normalizer $\mathfrak{N}_{O(3)}(\mathcal{S}_0)$ at $\vec r_0$, and $\mathfrak{Z}_g(\vec r)$ is a site-dependent operation valued in the centralizer $\mathfrak{Z}_{SO(3)}(\mathcal{S}_0)$.

We now show that, analogous to the fact that the two spin operations \( R_x(0)=E \) and \( R_x(\pi) \) fully determine conventional SPGs, the sector \( R_g^0 \) determines all subgroups \( \mathcal{G}\subset O(3)\times\operatorname{Isom}(\mathbb{R}^3) \) satisfying \( \mathcal{G}/\mathcal{S}_0\cong G_{\mathcal L} \). Since any two elements \( g,\widetilde{g}\in\mathcal{G} \) with the same lattice sector \( l_g=l_{\widetilde{g}} \) belong to the same coset of \( \mathcal{G}/\mathcal{S}_0 \), we have \( \widetilde{g}=s_{\widetilde{g}}g \) with \( s_{\widetilde{g}}\in\mathcal{S}_0 \). According to the above argument, we decompose \( g \) and \( \widetilde{g} \) as \( g=\big(\mathfrak{Z}_g(\vec r)\,R_g^0\mid\mid l_g\big) \) and \( \widetilde{g}=\big(\mathfrak{Z}_{\widetilde{g}}(\vec r)\,R_{\widetilde{g}}^0\mid\mid l_{\widetilde{g}}\big) \). Since \( \mathfrak{Z}_g(\vec r) \) commutes with \( \mathcal{S}_0 \), we obtain
\begin{align}
\widetilde{g}
=\big(\mathfrak{Z}_{\widetilde{g}}(\vec r)\,R_{\widetilde{g}}^0\mid\mid l_{\widetilde{g}}\big)
=\big(\mathfrak{Z}_{g}(\vec r)\,s_{\widetilde{g}}R_g^0\mid\mid l_{g}\big).
\end{align}
This implies that any two elements \( g,\widetilde{g}\in\mathcal{G} \) with the same lattice sector \( l_g=l_{\widetilde{g}} \) have the same normalizer sector modulo \( \mathcal{S}_0 \).  Consequently, there exists a map \( \hat{R}: l_g \mapsto R_g^0\mathcal{S}_0 \). We therefore can rewrite the coset representative \( R_g^0 \) as \( R_{l_g}^0 \) and \( \mathfrak{Z}_{\widetilde{g}} \) as \( \mathfrak{Z}_{l_g} \), emphasizing that they depend only on the lattice sector \( l_g \). Notably, since the centralizer group \( \mathfrak{Z}_{SO(3)}(\mathcal{S}_0) \) is a normal subgroup of the normalizer group \( \mathfrak{N}_{O(3)}(\mathcal{S}_0) \), the map \( \hat{R} \) is a group homomorphism
\begin{align}\label{GH}
\hat{R}:G_{\mathcal L}\to\mathfrak{N}_{O(3)}(\mathcal{S}_0)/\mathcal{S}_0 .
\end{align}
This homomorphism maps unitary (antiunitary) elements of $G_{\mathcal L}$ to cosets with determinant $+1$ ($-1$), and is well defined since $\mathcal{S}_0\subset SO(3)$.

Conversely, any such group homomorphism $\hat{R}$ uniquely specifies a subgroup of $O(3)\times \operatorname{Isom}(\mathbb{R}^3)$. This subgroup is constructed by assembling the cosets $(\hat{R}(l_g)\mid\mid l_g)$ for all $l_g\in G_{\mathcal L}$, with multiplication: 
\begin{small}
\begin{align}\label{DPM}
\left(s_{m}{R}^0_{l_{g_1}}\!\!\mid\mid l_{g_1}\right)\!\!\left(s_{n}R^0_{l_{g_2}}\!\!\mid\mid l_{g_2}\right)\!\!=\!\!\left(s_{m}R^0_{l_{g_1}}s_{n}R^0_{l_{g_2}}\!\!\mid\mid l_{g_1}l_{g_2}\right),
\end{align}\end{small}where $s_{m,n}\in\mathcal{S}_0$, and $R^0_{l_{g}}$ denotes a fixed coset representative of $\hat{R}(l_{g})$. 
A reformulation of conventional SSGs in this language is presented in the SM.

Each tSSG is formally similar to a subgroup of \( O(3)\times\operatorname{Isom}(\mathbb{R}^3) \), as illustrated by the relation between \( \mathcal G_3 \) and \( \mathcal G_2 \) in Eqs.~(\ref{GroupStructurescrC4}) and~(\ref{GroupStructuretildeC4}). These `similar' groups share the same group action \( \phi : G_{\mathcal L} \to \operatorname{Aut}(\mathcal Z(\mathcal S_0)) \), induced by conjugation of the normalizer sector on the center of \( \mathcal{S}_0 \), and are distinguished by elements of the second group cohomology $ H^{2}_{\phi}(G_{\mathcal L}, \mathcal Z(\mathcal S_0))$. Indeed, the centralizer sector $\mathfrak Z_{l_g}(\vec r)$ encodes these cohomology invariants and gives rise to tSSGs that cannot be embedded into $O(3)\times\operatorname{Isom}(\mathbb{R}^3)$. When $\mathfrak{Z}_{l_g}(\vec r)$ is present, the multiplication between elements
$\widetilde{g}_1=\big(\mathfrak{Z}_{l_{g_1}}(\vec r)\,s_{\widetilde{g}_1}R_{l_{g_1}}^0 \mid\mid l_{g_1}\big)$ and
$\widetilde{g}_2=\big(\mathfrak{Z}_{l_{g_2}}(\vec r)\,s_{\widetilde{g}_2}R_{l_{g_2}}^0 \mid\mid l_{g_2}\big)$
takes the form (see the SM)
\begin{align}\label{tSSGs-mul}
\widetilde{g}_1\cdot \widetilde{g}_2=\omega_2(l_{g_1},l_{g_2})\; \widetilde{g_1g_2}. 
\end{align}
where $\widetilde{g_1g_2}=(\mathfrak{Z}_{l_{g_1}l_{g_2}}(\vec r)\,s_{\widetilde{g}_1}R_{l_{g_1}}^0s_{\widetilde{g}_2}R_{l_{g_2}}^0\mid\mid l_{g_1}l_{g_2})$, and $\omega_2(l_{g_1},l_{g_2})\in \mathcal{Z}(\mathcal S_0)$ is a two-cocycle. For $\mathfrak{Z}_{{l_g}}(\vec r)\equiv E$, the product (\ref{tSSGs-mul}) is reduced to the law (\ref{DPM}).

\begin{figure}[t]
  \centering
\includegraphics[scale=0.45]{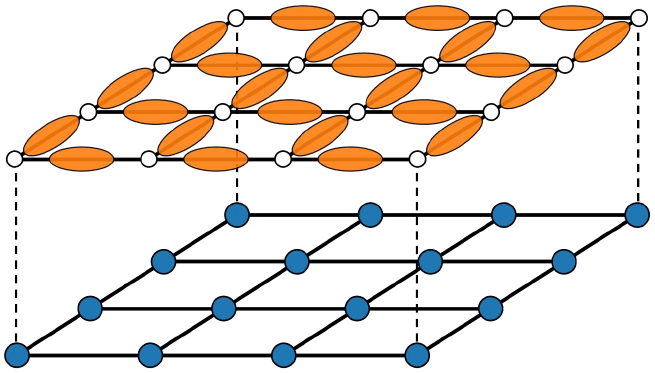}
\caption{The Hamiltonian~(\ref{CoexistenceHam}) realizing tSSGs with arbitrary $\mathcal{S}_0$. The upper layer represents the spin-exchange Hamiltonian $H_{\mathrm{bond}}$, while the lower layer represents the on-site single-ion anisotropy Hamiltonian $H_{\mathrm{site}}$.}\label{shuangcheng}
\end{figure}

We now present a universal construction of models realizing tSSGs. This mechanism is rigorously supported by the cocycle free construction of group extensions~\cite{Serre:finite-groups}. In general, tSSGs with  any given $\mathcal S_0$ can arise from the coexistence of on-site single-ion anisotropy Hamiltonians $H_{\mathrm{site}}$, with spin-only group $\mathcal S_0$, and bond spin interactions $H_{\mathrm{bond}}$, whose spin-only group is $\mathcal Z(\mathcal S_0)$, see Fig.~\ref{shuangcheng}. In this setting, the group homomorphism $\hat{R}$ is fixed by the symmetries of $H_{\mathrm{site}}$, while $H_{\mathrm{bond}}$ admits site-dependent sector valued in
$\mathfrak{Z}_{SO(3)}(\mathcal{S}_0) \subseteq \mathfrak{Z}_{SO(3)}\!\big(\mathcal Z(\mathcal S_0)\big)$.
A Hamiltonian with such tSSG symmetry therefore takes the generic form
\begin{align}\label{CoexistenceHam}
H=\lambda_1 H_{\mathrm{site}}+\lambda_2 H_{\mathrm{bond}}, \qquad
\lambda_1,\lambda_2\in\mathbb{R}.
\end{align}
In the limit $\lambda_2 \to 0$, the model reduces to one whose symmetries form an SSG with spin-only group $\mathcal{S}_0$ and no centralizer sector. In contrast, for $\lambda_1 \to 0$, it reduces to a model whose symmetries form a tSSG with spin-only group $\mathcal{Z}(\mathcal{S}_0)$, with the centralizer sector taking values in $\mathfrak{Z}_{SO(3)}(\mathcal{S}_0)$. Rigorous proofs are provided in the SM. 

\paragraph*{Topological quadrupolar bands on spin Brillouin Klein bottle.—}
As an application of the tSSG theory, we show that spin models protected by tSSGs can yield quadrupolar excitations on a spin Brillouin zone whose fundamental domain is topologically a Klein bottle, arising from the presence of a free action, namely a glide reflection in the spin Brillouin zone~\cite{chen2022brillouin}. This Klein-bottle topology reshapes the topological classification of quadrupolar excitation bands~\cite{segal1968equivariant,note2}: every topological band protected by the glide-reflection symmetry is topologically equivalent to a conventional topological band defined on the Klein bottle. We refer to these excitations as topological quadrupolars on spin Brillouin Klein bottle. Following Ref.~\cite{chen2022brillouin}, they are classified by a \(\mathbb{Z}_2\) invariant, rather than the $\mathbb{Z}$-valued Chern invariant on the Brillouin torus.

Consider the following anisotropic spin-1 Hamiltonian with $\mathcal{S}_0=\mathcal{C}_2$ on the lattice $\vec{r}=(r_x,r_y)\in\mathbb{Z}\times\mathbb{Z}$ with wallpaper group $pm$:  
\begin{equation}\label{quadrupolarH}
H=\sum_{\vec r}
\Big[
\bm S_{\vec r}^{\mathsf T}{\bm J}_{1}\bm S_{\vec r+\hat x}
+
\bm S_{\vec r}^{\mathsf T}{\bm{J}}_{2}\bm S_{\vec r+\hat y}
\Big]
+
D\sum_{\vec r}(S_{\vec r}^z)^2,
\end{equation}
where $\hat x=(1,0)$, $\hat y=(0,1)$, $ \bm{S}_{\vec{r}} = (S^x_{\vec{r}}, S^y_{\vec{r}}, S^z_{\vec{r}})^T $, and \begin{small}\begin{align}
{\bm J}_{1}=
\begin{pmatrix}
a+b &\lambda & 0\\
-\lambda & a-b & 0\\
0 & 0 & c
\end{pmatrix},\quad {\bm J}_{2}=
\begin{pmatrix}
0 & \beta+\gamma & 0\\
\beta-\gamma &0 & 0\\
0 & 0 & \alpha
\end{pmatrix}\notag
\end{align}\end{small}with $a,b,c,\lambda,\alpha,\beta,\gamma,D$ all arbitrary real parameters. The tSSG symmetries of the Hamiltonian (\ref{quadrupolarH}) are the following: $\widetilde{T}_x=\left(E\mid\mid\mathbb{T}_x\right)$, $\widetilde{T}_y=(E\mid\mid\mathbb{T}_y)$, \( \widetilde{\mathcal{M}}_x = (\mathfrak{Z}_{\mathcal{M}_x}(\vec{r})R_{[100]}(\pi)\mid\mid \mathcal{M}_x) \) with $\mathfrak{Z}_{\mathcal{M}_x}(\vec{r})=(R_{[001]}(\pi))^{r_y}$. Notice that $\widetilde{M}_x\widetilde{T}_y\widetilde{M}_x^{-1}\widetilde{T}_y^{-1}=R_{[001]}(\pi)\in\mathcal{S}_0.$ The model is analyzed in detail in the SM using linear flavor wave theory, and we present the main results below.

In the Cartesian spin-1 basis $\{|x\rangle,|y\rangle,|z\rangle\}$, the large $D$ single-ion anisotropy selects the ferroquadrupolar vacuum $|z\rangle$ at every site. Condensing the $|z\rangle$ flavor and keeping harmonic terms gives $Q_{xz}\sim-(b_x+b_x^{\dagger})$ and $Q_{yz}\sim-(b_y+b_y^{\dagger})$, so the $\{Q_{xz},Q_{yz}\}$ sector is a genuine one-particle bosonic problem. Thus, one can write down a four-band bosonic Bogoliubov-de Gennes Hamiltonian in the basis $(b_{x,\bm k}, b_{y,\bm k}, b^{\dagger}_{x,-\bm k}, b^{\dagger}_{y,-\bm k})^{\mathrm{T}}$. Meanwhile, it is shown that the harmonic kernel equivalently reduces to a $2\times2$ Hermitian matrix,
\begin{equation}
\begin{aligned}
M_1(\bm k)=&\,D\,\mathbb{I}_2+4a\cos k_x\,\mathbb{I}_2-4\beta\cos k_y\,\sigma_x \\
&+4(\lambda\sin k_x+\gamma\sin k_y)\sigma_y-4b\cos k_x\,\sigma_z,
\end{aligned}
\label{eq:M1letter}
\end{equation}
whose two positive bosonic branches are
$\omega_{\pm}(\bm k)=\sqrt{D[D+4a\cos k_x\pm4\rho(\bm k)]}$,
with $\rho^2=b^2\cos^2k_x+\beta^2\cos^2k_y+(\lambda\sin k_x+\gamma\sin k_y)^2$. The crucial point is 
\begin{equation}
\widetilde{\mathcal{M}}_x:\quad {\bm b}_{(k_x,k_y)}
\mapsto e^{-\mi k_x}(-\sigma_z){\bm b}_{(-k_x,\,k_y+\pi)}, 
\label{eq:Mx-kspace-action}
\end{equation} which is a momentum glide reflection. Indeed, the kernel itself also obeys the glide-reflection sewing relation:
\begin{equation}
M_1(-k_x,k_y+\pi)=(-\sigma_z) M_1(k_x,k_y)(-\sigma_z)^{-1},
\label{eq:glideletter}
\end{equation}
The bosonic spectrum therefore satisfies $\omega_{\pm}(k_x,k_y)=\omega_{\pm}(-k_x,k_y+\pi)$, and the fundamental domain of the sector is a Klein bottle rather than a torus~\cite{chen2022brillouin}.

This bulk glide structure has a sharp boundary consequence. For open boundaries along $x$, the exactly solvable line $a=0$, $b=\lambda$ admits closed-form edge modes with energies
$\omega_{L,R}(k_y)=\sqrt{D(D\pm4\beta\cos k_y)}$ and decay factor
$r_{L,R}(k_y)=\mp\mi\gamma\sin k_y/\lambda$. The left and right edge branches are finally sewn by nonlocal twist $\omega_L(k_y)=\omega_R(k_y+\pi)$. Their existence is controlled by the winding of the off-diagonal polynomial
$q(k_x;k_y)=-4(\lambda e^{\mi k_x}+\mi\gamma\sin k_y)$,  namely\\
\begin{small}\begin{equation}
\nu(k_y)=\frac{1}{2\pi\mi}\int_{-\pi}^{\pi} \! \mathrm{d}k_x\,\partial_{k_x}\log q
=
\begin{cases}
1,& |\gamma\sin k_y|<|\lambda|,\\
0,& |\gamma\sin k_y|>|\lambda|,
\end{cases}\notag
\end{equation}\end{small}\\
which counts the number of roots of $q(z;k_y)$ inside the unit disk.
This $\mathbb{Z}_2$ quantity is the slice invariant governing edge localization: any symmetry-preserving perturbation that leaves the glide sewing intact and keeps $q$ nonzero on the unit circle cannot change $\nu(k_y)$ and hence cannot remove the corresponding boundary mode.
For the lower positive-frequency band, the bulk Klein-bottle invariant satisfies $\nu_{\mathrm K}=\nu(-\pi/2)\pmod2$. For instance, setting $\lambda=1$ and $\gamma=0.9$ gives $\nu(k_y)=1$ for every $k_y$, implying both $\nu_{\mathrm K}=1$ and a nonlocally sewn edge pair spanning the entire spin Brillouin zone rather than fragmentary edge arcs.

\paragraph*{Discussion and Outlook.—}
Physically, since the homomorphism $\hat{R}$ defined in Eq.~\eqref{GH} characterizes all subgroups of the symmetry group of the Heisenberg model, it encodes every symmetry-breaking pattern induced by anisotropic perturbations in realistic materials. On the other hand, since DM vectors act as a gauge field for spin currents~\cite{katsura2005spin}, the site-dependent centralizer sector $\mathfrak{Z}_{g}(\vec{r})$ in Fig.~\ref{model-tSSG} admits a natural interpretation as a \emph{spin gauge}. From this perspective, tSSGs with nontrivial $\mathfrak{Z}_{g}(\vec{r})$ arise from gauging the centralizer of $\mathcal{S}_0$.

Both $\hat{R}$ and gauge-dressed spin symmetries can break the homomorphic relation between lattice and spin operations (see  Sec.3 of the SM), thereby generating magnetic excitations protected by \emph{projective} crystalline symmetries. These quasiparticles can exhibit Brillouin platycosm topology~\cite{ZhangYangZhao2025ProjectiveCrystalSymmetry,zhang2025brillouin}. To our knowledge, this provides the first interacting spin model supporting topological magnetic excitations  on spin Brillouin Klein bottle.  This work motivates the theoretical and experimental search for magnetic excitations on Brillouin platycosms, in real magnetic materials and on programmable platforms for the quantum simulation of magnetism alike.

\paragraph*{Acknowledgments.—} I am grateful to Changle Liu for introducing the flavor wave formalism, for assistance with numerical analysis of model (\ref{The-model}) and (\ref{quadrupolarH}), as well as for valuable discussions. I also thank Zheng-Xin Liu, Ziyin Song and Chen Zhang for valuable comments and discussions. Part of this work, including the early development of model (\ref{The-model}), was carried out during a visit to Renmin University of China, with support from Zheng-Xin Liu. This work was supported by the General Research Fund of Hong Kong (Nos. 17301224 and 17302525).

\bibliography{tSSGs_references}

\begin{thebibliography}{64}
\bibitem[{not({\natexlab{a}})}]{note1}
\bibinfo{note}{From the perspective of group extensions,
  $\mathbb{Z}_4=\{1,z,z^2,z^3\}$ admits two distinct actions on $SO(2)$,
  corresponding to the trivial homomorphism $\phi_0:\mathbb{Z}_4 \to
  \operatorname{Aut}(SO(2)) \cong \mathbb{Z}_2$ and the nontrivial one
  $\phi_1:\mathbb{Z}_4 \to \operatorname{Aut}(SO(2))$. For $\phi_0$, the unique
  solution is the direct product $SO(2)\times\mathbb{Z}_4$. For $\phi_1$, there
  exist two inequivalent solutions: the semidirect product
  $SO(2)\rtimes\mathbb{Z}_4$, and a twisted extension characterized by
  $(z)^4=-E_2\in SO(2)$.}
\bibitem[{not({\natexlab{b}})}]{note2}
\bibinfo{note}{Mathematically, an equivariant bundle with structure group \( G
  \) acting freely on a base space \( X \) is equivalent to a bundle over the
  orbit space \( X/G \).}
\end{thebibliography}

\end{document}


\title{Supplemental Material for ``On the Symmetries of Anisotropic Spin Interaction Models''}

\author{Arist Zhenyuan Yang}
\email{zhenyuan@hku.hk}
\affiliation{Department of Physics and HK Institute of Quantum Science \& Technology, The University of Hong Kong, Pokfulam Road, Hong Kong, China}

\date{\today}

\maketitle
\tableofcontents

\setcounter{section}{0}
\setcounter{subsection}{0}
\setcounter{equation}{0}
\setcounter{figure}{0}
\setcounter{table}{0}

\section{Reformulation of conventional SSGs}

Using the formulation developed in the main text, conventional SSGs are readily recovered. For classical magnetic orders, the unitary spin-only group $\mathcal{S}_0$ falls into three cases: $SO(3)$, $SO(2)_z$, and the trivial group $\mathcal{C}_1$, corresponding to nonmagnetic systems, collinear magnetic order, and coplanar or noncoplanar magnetic order, respectively. The latter two are distinguished by the presence or absence of an effective time-reversal symmetry. Nonmagnetic systems require $( -E \mid\mid \mathcal{T} )$ to be a symmetry, whereas collinear magnetic order requires $( -R_x(\pi) \mid\mid \mathcal{T} )$ to be a symmetry.

The classification of SSGs then reduces to computing the group homomorphisms
\begin{align}
\hat{R}: G_{\mathcal{L}} \to \mathfrak{N}_{O(3)}(\mathcal{S}_0)/\mathcal{S}_0 ,
\end{align}
subject to the constraint that antiunitary elements are mapped to spin operations with determinant $-1$.
The different cases are discussed separately below.

\paragraph*{1. Nonmagnetic systems.} When $\mathcal{S}_0 = SO(3)$, the center is trivial, $\mathcal{Z}(SO(3))=\mathcal{C}_1$, and
\[
\mathfrak{N}_{O(3)}(SO(3))/SO(3)=\mathbb{Z}_2^{-}=\{E,-E\}.
\]
Accordingly, the group homomorphism $\hat{R}$ admits two possibilities, depending on whether $G_{\mathcal{L}}$ contains antiunitary elements. If $G_{\mathcal{L}}$ is a space group, $\hat{R}$ is trivial. If $G_{\mathcal{L}}$ is a magnetic space group, $\hat{R}$ is uniquely fixed by requiring its kernel to be the maximal unitary subgroup of $G_{\mathcal{L}}$. For a given $\hat{R}$, the corresponding SSG $\mathcal{G}$ is given by
\[
\mathcal{G}=\{(SO(3)\hat{R}(g)\mid\mid g)\;|\; g\in G_{\mathcal{L}}\}.
\]

In this case, the presence of the symmetry $(-E\mid\mid\mathcal{T})$ implies
$\mathcal{G}\cong O(3)\times SG$.

\paragraph*{2. Collinear magnetic order.}
When $\mathcal{S}_0 = SO(2)_z$, the normalizer quotient is
\begin{equation}
\mathfrak{N}_{O(3)}(SO(2)_z)/SO(2)_z
= \mathbb{Z}_2^{x}\times \mathbb{Z}_2^{-}
= \{E,\, R_x(\pi),\, -E,\, -R_x(\pi)\}.
\end{equation}
Accordingly, the group homomorphism $\hat{R}$ decomposes as follows: unitary elements of $G_{\mathcal{L}}$ are mapped to $\{E, R_x(\pi)\}$, while antiunitary elements are mapped to $\{-E, -R_x(\pi)\}$.

For collinear magnetic order, the effective time-reversal symmetry is fixed as $\hat{R}(\mathcal{T})=-R_x(\pi)$. The classification of $\hat{R}$ therefore reduces to that of one-dimensional real representations of the maximal unitary subgroup of $G_{\mathcal{L}}$, equivalently, of a crystallographic space group.

\paragraph*{3. Coplanar magnetic order.}
We assume coplanar spin moments lying in the $x$--$y$ plane. In this case,
\begin{equation}
\mathfrak{N}_{O(3)}(\mathcal{C}_1)/\mathcal{C}_1=O(3).
\end{equation}
Coplanar systems admit the effective time-reversal symmetry $(-R_z(\pi)\mid\mid\mathcal{T})$, which fixes $\hat{R}(\mathcal{T})=-R_z(\pi)$. Since the spin-only group is trivial, this effective time-reversal operation commutes with all symmetry operations. This implies that the image of $\hat{R}$ restricted to the maximal unitary subgroup of $G_{\mathcal{L}}$ lies in $\operatorname{diag}[g,\det(g)]\subset SO(3)$ with $g\in O(2)_z$. Consequently, the classification reduces to enumerating the two-dimensional {\it real} representations of the maximal unitary subgroup of $G_{\mathcal{L}}$.

\paragraph*{4. Noncoplanar magnetic order.}
For noncoplanar magnetic order, the spin-only group is trivial, $\mathcal{S}_0=E$, and no effective time-reversal symmetry is present. Consequently, \(G{\mathcal{L}}\) cannot be a type-II magnetic space group, and must instead be an MSG of type I, III, or IV. Since
\begin{equation}
\mathfrak{N}_{O(3)}(\mathcal{C}_1)/\mathcal{C}_1 = O(3),
\end{equation}
the corresponding homomorphism $\hat{R}$ is specified by a three-dimensional real orthogonal representation of the underlying ordinary space group of $G_{\mathcal{L}}$.

We conclude this section by addressing symmetry twists in conventional SSGs. Since neither $SO(3)$ nor the trivial group $\mathcal{C}_1$ has a nontrivial center, the associated second cohomology groups are trivial. Consequently, twisted SSGs with nontrivial cohomology invariants can arise only in collinear magnetic systems, as illustrated by the examples discussed in the main text.

\section{The derivation of the group multiplication law for twisted SSGs}

As shown in the main text, each element $g$ of a twisted SSG $\mathcal{G}$ admits a decomposition into a lattice sector $l_g$, a normalizer sector $R_g^0$ at a reference site $\vec{r}_0$, and a site-dependent centralizer sector $\mathfrak{Z}_g(\vec r)$, namely $ g=\big(\mathfrak{Z}_g(\vec r)\,R_g^0 \mid\mid l_g\big)$. Furthermore, for two $g,\widetilde{g}$ lying in the same coset $\mathcal{G}/\mathcal{S}_0$, we can write $g=\big(\mathfrak{Z}_{l_g}(\vec r)\,R_{l_g}^0 \mid\mid l_g\big)$ and $\widetilde{g}=\big(\mathfrak{Z}_{l_{g}}(\vec r)\,s_{\widetilde{g}}R_{l_g}^0 \mid\mid l_g\big)$ with $s_{\widetilde{g}}\in\mathcal{S}_0$. We now show that the multiplication between elements $\widetilde{g}_1=\big(\mathfrak{Z}_{l_{g_1}}(\vec r)\,s_{\widetilde{g}_1}R_{l_{g_1}}^0 \mid\mid l_{g_1}\big)$ and $\widetilde{g}_2=\big(\mathfrak{Z}_{l_{g_2}}(\vec r)\,s_{\widetilde{g}_2}R_{l_{g_2}}^0 \mid\mid l_{g_2}\big)$ takes the form
\begin{align}\label{SM:tSSGs-mul}
\widetilde{g}_1\cdot \widetilde{g}_2=\omega_2(l_{g_1},l_{g_2})\; \widetilde{g_1g_2}. 
\end{align}
where $\widetilde{g_1g_2}=(\mathfrak{Z}_{l_{g_1}l_{g_2}}(\vec r)\,s_{\widetilde{g}_1}R_{l_{g_1}}^0s_{\widetilde{g}_2}R_{l_{g_2}}^0\mid\mid l_{g_1}l_{g_2})$, and $\omega_2(l_{g_1},l_{g_2})\in \mathcal{Z}(\mathcal S_0)$ is a two-cocycle.

The result follows from a direct calculation:
\begin{align}
\widetilde{g}_1\cdot \widetilde{g}_2=&\left(\mathfrak{Z}_{l_{g_1}}(\vec r)\,s_{\widetilde{g}_1}R_{l_{g_1}}^0 \mid\mid l_{g_1}\right) \left(\mathfrak{Z}_{l_{g_2}}(\vec r)\,s_{\widetilde{g}_2}R_{l_{g_2}}^0 \mid\mid l_{g_2}\right)\notag\\
=&\left( \mathfrak{Z}_{l_{g_1}}(\vec r)\,s_{\widetilde{g}_1}R_{l_{g_1}}^0 l_{g_1} \mathfrak{Z}_{l_{g_2}}(\vec r)\,s_{\widetilde{g}_2}R_{l_{g_2}}^0 \mid\mid  l_{g_2} \right)\notag\\
=& \left([ \mathfrak{Z}_{l_{g_1}}(\vec r)\,s_{\widetilde{g}_1}R_{l_{g_1}}^0]l_{g_1}[ \mathfrak{Z}_{l_{g_2}}(\vec r)\,s_{\widetilde{g}_2}R_{l_{g_2}}^0]l_{g_1}^{-1}   \mid\mid l_{g_1}l_{g_2} \right)\notag\\
=& \left([ \mathfrak{Z}_{l_{g_1}}(\vec r)\,s_{\widetilde{g}_1}R_{l_{g_1}}^0][ \mathfrak{Z}_{l_{g_2}}(l_{g_1}^{-1}\cdot\vec r)\,s_{\widetilde{g}_2}R_{l_{g_2}}^0]   \mid\mid l_{g_1}l_{g_2} \right)\notag\\
=&\left(      [ \mathfrak{Z}_{l_{g_1}}(\vec r)\,s_{\widetilde{g}_1}][R_{l_{g_1}}^0 \mathfrak{Z}_{l_{g_2}}(l_{g_1}^{-1}\cdot\vec r)(R_{l_{g_1}}^0)^{-1}][R_{l_{g_1}}^0s_{\widetilde{g}_2}R_{l_{g_2}}^0]   \mid\mid l_{g_1}l_{g_2}      \right)\notag\\
=&\left(       \mathfrak{Z}_{l_{g_1}}(\vec r)[R_{l_{g_1}}^0 \mathfrak{Z}_{l_{g_2}}(l_{g_1}^{-1}\cdot\vec r)(R_{l_{g_1}}^0)^{-1}]s_{\widetilde{g}_1}[R_{l_{g_1}}^0s_{\widetilde{g}_2}R_{l_{g_2}}^0]   \mid\mid l_{g_1}l_{g_2}      \right)\notag\\
=&\left(      \left[ \mathfrak{Z}_{l_{g_1}}(\vec r)R_{l_{g_1}}^0 \mathfrak{Z}_{l_{g_2}}(l_{g_1}^{-1}\cdot\vec r)(R_{l_{g_1}}^0)^{-1}\mathfrak{Z}_{l_{g_1}l_{g_2}}(\vec{r})^{-1}\right]\mathfrak{Z}_{l_{g_1}l_{g_2}}(\vec{r})[s_{\widetilde{g}_1}R_{l_{g_1}}^0s_{\widetilde{g}_2}R_{l_{g_2}}^0]   \mid\mid l_{g_1}l_{g_2}      \right)\notag\\
=& \left[ \mathfrak{Z}_{l_{g_1}}(\vec r)R_{l_{g_1}}^0 \mathfrak{Z}_{l_{g_2}}(l_{g_1}^{-1}\cdot\vec r)(R_{l_{g_1}}^0)^{-1}\mathfrak{Z}_{l_{g_1}l_{g_2}}(\vec{r})^{-1}\right]\left(  \mathfrak{Z}_{l_{g_1}l_{g_2}}(\vec{r})s_{\widetilde{g}_1}R_{l_{g_1}}^0s_{\widetilde{g}_2}R_{l_{g_2}}^0  \mid\mid l_{g_1}l_{g_2}      \right)\\
\equiv &\;\; \omega_2(l_{g_1},l_{g_2})\; \widetilde{g_1g_2},
\end{align}
where the fourth equality follows from the fact that $s_{\widetilde{g}_2} R_{l_{g_2}}^0$ is a global spin operation, and the sixth equality follows from the fact that the centralizer is a normal subgroup of the normalizer. The factor $\omega_2(l_{g_1},l_{g_2})$ is a product of centralizer elements and therefore commutes with $\mathcal{S}_0$. Moreover, since $\omega_2(l_{g_1},l_{g_2}) \in \mathcal{S}_0$, it is a global element of the center $\mathcal{Z}(\mathcal{S}_0)$. Furthermore, associativity of the group multiplication implies that $\omega_2$ satisfies the two-cocycle condition and hence defines an element of the second cohomology group $H^2_{\phi}(G_{\mathcal{L}}, \mathcal{Z}(\mathcal{S}_0))$. Here the group action $ \phi : G_{\mathcal L} \to \operatorname{Aut}(\mathcal Z(\mathcal S_0)) $ is induced by the conjugation action of the normalizer sector on $\mathcal{Z}(\mathcal{S}_0)$.

\section{Group Extensions with arbitrary kernels and an alternative construction of twisted SSGs}

Consider the following group extension 
\begin{align}\label{SM:GExt}
1\to \mathcal{S}_0\to\mathcal{G}\to G_{\mathcal{L}}\to 1.
\end{align}
Below, we present {\it three} complementary constructions for $\mathcal{G}$: one for abelian kernels $\mathcal{S}_0$ and two for nonabelian kernels $\mathcal{S}_0$. The latter two consist of a cocycle-dependent construction, which directly leads to the formulation of twisted SSGs in the main text, and a cocycle-free construction, which proves that twisted SSGs with nonabelian spin-only groups arise generically from the coexistence of on-site single-ion anisotropy and bond spin-exchange interactions. For a standard introduction to group extension theory, see Serre's book~\cite{Serre:finite-groups}.

We begin with the case in which the spin-only group $\mathcal{S}_0$ is abelian.
A group homomorphism $ \hat{R}: G_{\mathcal{L}} \to \mathfrak{N}_{O(3)}(\mathcal{S}_0)/\mathcal{S}_0 $ uniquely specifies an action $$ \phi: G_{\mathcal{L}} \to \operatorname{Aut}(\mathcal{S}_0).$$ 
This action $\phi$ in turn determines the cohomology groups $H^n_{\phi}(G_{\mathcal{L}}, \mathcal{S}_0)$. The isomorphism classes of $\mathcal{G}$ in Eq.~(\ref{SM:GExt}) are in bijection with the elements of the second cohomology group $H^2_{\phi}(G_{\mathcal{L}}, \mathcal{S}_0)$. 
In particular, let $\omega_2$ be a normalized two-cocycle representing a class $ [\omega_2]\in H^2_{\phi}(G_{\mathcal L},\mathcal S_0)$. Then one can construct a group $\mathcal G$ whose underlying set is $\mathcal S_0\times G_{\mathcal L}$, equipped with the multiplication law
\begin{align}\label{SM:abelianG}
(s_1,l_1)\diamond (s_2,l_2) = \big(s_1\,\phi(l_1)(s_2)\,\omega_2(l_1,l_2),\; l_1l_2\big),
\end{align}
for $s_{1,2}\in\mathcal S_0$ and $l_{1,2}\in G_{\mathcal L}$. 
Notice that the homomorphism $\hat{R}$ determines a two-cocycle
\begin{align}\label{R-cocycle}
\omega_{2}^{\hat{R}}(l_1,l_2)=R_{l_{1}}R_{l_{2}}R_{l_{1}l_{2}}^{-1},
\end{align}
with $R_{l_g}$ a section of $\hat{R}$. Since the action $\phi:G_{\mathcal L}\to\operatorname{Aut}(\mathcal S_0)$ is fixed by the normalizer sector $R_{G_{\mathcal{L}}}$, we can rewrite (\ref{SM:abelianG}) by explicitly attaching the normalizer sector  $R_{l_g}$ to lattice sector $l_g$, namely
\begin{align}\label{SM:s9s11}
 \text{(\ref{SM:abelianG})}\;\;\Longleftrightarrow\;\; (s_1R_{l_1}, l_1)\diamond (s_2R_{l_2}, l_2) =\big(s_1\,R_{l_{1}}s_2R_{l_{1}}^{-1}\,\omega_2(l_1,l_2) R_{l_1l_2},\; l_1l_2\big),
\end{align}
Remarkably, choosing the cocycle $\omega_2$ in Eq.~(\ref{SM:s9s11}) to be $\omega_2^{\hat R}$ defined in Eq.~(\ref{R-cocycle}), the multiplication law reduces to
\begin{align}\label{S12}
(s_1 R_{l_1}, l_1)\cdot(s_2 R_{l_2}, l_2)=\big( s_1\, R_{l_1} s_2 R_{l_2} R_{l_1 l_2}^{-1}R_{l_1 l_2},\, l_1 l_2 \big) = \big( s_1\, R_{l_1} s_2 R_{l_2}, \;l_1 l_2
\big),
\end{align}
which is precisely the multiplication rule of a subgroup of the direct product group $O(3)\times \operatorname{Isom}(\mathbb{R}^3)$.
Hence, in comparison with Eq.~(\ref{SM:abelianG}), the multiplication law of twisted SSGs can be written in the abstract form
\begin{align}\label{AAF}
(s_1 R_{l_1}, l_1)\diamond(s_2 R_{l_2}, l_2)=\widetilde{\omega}_2(l_1,l_2)\big( s_1\, R_{l_1} s_2 R_{l_2}, \;l_1 l_2 \big),\text{ with }\widetilde{\omega}_2=\omega_2\cdot\left(\omega_2^{\hat{R}}\right)^{-1}. 
\end{align}
From a physical perspective, this additional cocycle $\widetilde{\omega}_2$ is realized by site-dependent centralizer sector $\mathfrak{Z}_{g}(\vec{r})$.

It is important to note that the triviality of the two-cocycle
$\omega_{2}^{\hat{R}}(l_1,l_2)$ is equivalent to the statement that the
correspondence between a lattice operation $l$ and its associated spin
operation $R_l$ is a group homomorphism. Equivalently, the subgroup of
$O(3)\times \operatorname{Isom}(\mathbb{R}^3)$ defined in
Eq.~\eqref{S12} is the graph of an $O(3)$-valued representation of
$G_{\mathcal{L}}$, as in conventional SSGs. In this case, the subgroup
can be written as the semidirect product
$\mathcal{S}_0\rtimes G_{\mathcal{L}}$. After absorbing the spin
operations into the magnetic-excitation basis, these excitations carry
a linear representation of $G_{\mathcal{L}}$. By contrast, when
$\omega_{2}^{\hat{R}}(l_1,l_2)$ is nontrivial, magnetic excitations are
generally protected by projective crystalline symmetry.

We now turn to the case of nonabelian spin-only groups $\mathcal S_0$.  The overall structure parallels the abelian case, but with important modifications. 
A group homomorphism $ \hat{R}: G_{\mathcal L} \to \mathfrak{N}_{O(3)}(\mathcal S_0)/\mathcal S_0 $
uniquely induces an action
$$
\phi: G_{\mathcal L} \to \operatorname{Aut}\!\big(\mathcal Z(\mathcal S_0)\big),
$$
and hence defines the cohomology groups $H^n_{\phi}\!\big(G_{\mathcal L}, \mathcal Z(\mathcal S_0)\big)$. In contrast to the abelian case, the isomorphism classes of group extensions are no longer in one-to-one correspondence with elements of the second cohomology group $H^2_{\phi}\!\big(G_{\mathcal L}, \mathcal Z(\mathcal S_0)\big)$. Instead, this cohomology group characterizes the relative differences between extensions. Specifically, let $\mathcal{G}_0$ be an extension whose inner automorphisms induces the group action $\phi: G_{\mathcal L} \to \operatorname{Aut}\!\big(\mathcal Z(\mathcal S_0)\big)$. Then, one can construct all other extensions by revising the multiplication of $\mathcal{G}_0$ in the following way: choose a normalized two-cocycle $\omega_2$ representing a class in $H^2_{\phi}\!\big(G_{\mathcal L}, \mathcal Z(\mathcal S_0)\big)$, and take two elements $g_1, g_2 \in \mathcal G_0$. We define a new multiplication law by
\begin{align}\label{NAAL}
g_1 \diamond g_2
= \omega_2(l_{g_1}, l_{g_2})\, (g_1 \cdot g_2),
\end{align}
where $\cdot$ denotes the multiplication in $\mathcal G_0$. In our setting, $\mathcal G_0$ is generated by the group homomorphism $\hat R$, namely it is a subgroup of $O(3)\times \operatorname{Isom}(\mathbb{R}^3)$. 
Then (\ref{NAAL}) can be written as
\begin{align}
(s_1 R_{l_1}, l_1)\diamond(s_2 R_{l_2} ,l_2)=\omega_2(l_1,l_2)\big( s_1\, R_{l_1} s_2 R_{l_2}, \;l_1 l_2 \big).
\end{align}
This construction renders the formal structure of nonabelian group extensions identical to that of the abelian case in Eq.~(\ref{AAF}). For nonabelian $\mathcal S_0$, the cocycle $\omega_2$ is likewise realized by the site-dependent centralizer sector $\mathfrak Z_g(\vec r)$.

The construction above is cocycle-dependent. We now present a complementary cocycle-free construction of nonabelian group extensions. This approach is based on two group extensions with kernels $\mathcal{S}_0$ and $\mathcal{Z}(\mathcal{S}_0)$, respectively. We keep $\mathcal{G}_0$ as defined above, and let $\mathcal{F}$ denote a group extension of $G_{\mathcal{L}}$ by $\mathcal{Z}(\mathcal{S}_0)$. Consider the diagonal subgroup $\Delta_{ G_{\mathcal L}}\subset G_{\mathcal L}\times G_{\mathcal L}$, and define $ \mathcal G_0\times_{G_{\mathcal L}}\mathcal F $ be the inverse image of $\Delta_{ G_{\mathcal L}}$ under the map $G_0\times\mathcal{F}\to  G_{\mathcal L}\times  G_{\mathcal L}$. The group $\mathcal G_0\times_{G_{\mathcal L}}\mathcal F $ is called the product of $\mathcal{G}_0$ and $\mathcal{F}$ over $ G_{\mathcal L}$. It fits into the exact sequence
$$
1\to \mathcal S_0\times \mathcal Z(\mathcal S_0)
\to \mathcal G_0\times_{G_{\mathcal L}}\mathcal F
\to G_{\mathcal L}\to 1 .
$$
Let $ \Delta^-=\{(s,s^{-1})\in \mathcal S_0\times \mathcal Z(\mathcal S_0)\}$. Then $\Delta^-$ is a normal subgroup, and
$(\mathcal S_0\times \mathcal Z(\mathcal S_0))/\Delta^-\cong \mathcal S_0$.
Define $ \mathcal G := (\mathcal G_0\times_{G_{\mathcal L}}\mathcal F)/\Delta^- $. This yields an extension
\begin{align}\label{thirdextension}
1\to \mathcal S_0\to \mathcal G\to G_{\mathcal L}\to 1,
\end{align}
which is isomorphic to the twisted SSG $\mathcal{G}$ constructed from the cocycle
$\omega_2\in H^2_\phi(G_{\mathcal L},\mathcal Z(\mathcal S_0))$.

\begin{figure}[h]
  \centering
\includegraphics[scale=0.7]{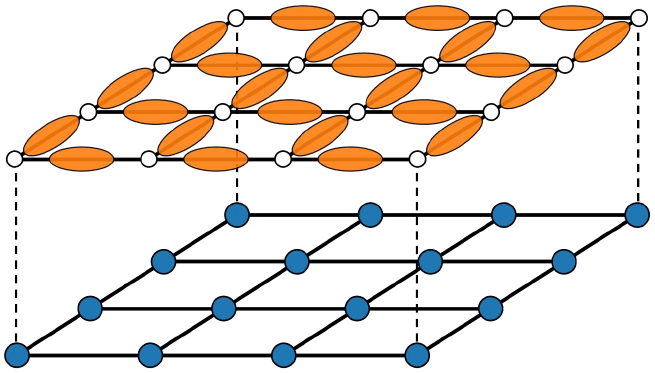}
\caption{Decomposition of the total Hamiltonian into on-site and bond interaction terms.}\label{SM:shuangcheng}
\end{figure} 

The cocycle-free construction has a direct physical application. 
It shows that twisted SSGs with nonabelian $\mathcal{S}_0$ arise generically from the coexistence of on-site single-ion anisotropy with spin-only group $\mathcal{S}_0$ and bond spin-exchange interactions with spin-only group $\mathcal{Z}(\mathcal{S}_0)$. In general, consider Hamiltonians of the form
$$
H = H_{\text{site}} + H_{\text{bond}},
$$
where the on-site term $H_{\text{site}}$ has spin-only group $\mathcal{S}_0$, while the bond term $H_{\text{bond}}$ has spin-only group $\mathcal{Z}(\mathcal{S}_0)$, as illustrated in Fig.~\ref{SM:shuangcheng}. Let $\mathcal{G}_0$ and $\mathcal{F}$ denote the symmetry groups of $H_{\text{site}}$ and $H_{\text{bond}}$, respectively. In the absence of additional constraints, the symmetry of the two seperated systems $H_{\text{site}}$ and $H_{\text{bond}}$ are the direct product $\mathcal{G}_0\times\mathcal{F}$, and the spin-only group is $\mathcal{S}_0\times\mathcal{Z}(\mathcal{S}_0)$. In order to obtain the symmetry group $\mathcal{G}$ of the full Hamiltonian $H$, two constraints must be imposed.  First, the two subsystems must share the same lattice operations.  Second, their on-site symmetry operations must coincide. Accordingly, the direct product group $\mathcal{G}_0 \times \mathcal{F}$ must be restricted to the fiber product $\mathcal{G}_0 \times_{G_{\mathcal{L}}} \mathcal{F}$, and on-site operations $(s,b) \in \mathcal{S}_0 \times \mathcal{Z}(\mathcal{S}_0)$ are required to be trivialized whenever $s \cdot b = 1$. The two constraints uniquely fix the symmetry group $\mathcal{G}$ of the full Hamiltonian $H = H_{\text{site}} + H_{\text{bond}}$. In particular, $\mathcal{G}$ coincides precisely with the group extension defined in Eq.~(\ref{thirdextension}).

\newpage

\section{Exact results of the spin-1 model}

We denote lattice sites by \( \br=(r_x,r_y)\in\mathbb{Z}\times\mathbb{Z} \), and write
\( \bm{S}_{\br}=(S^x_{\br},S^y_{\br},S^z_{\br})^T \).  The spin-1 anisotropic Hamiltonian is
\begin{equation}
H=\sum_{\br}\Big[\bm{S}_{\br}^{T}\cdot \bm{J}_1\cdot\bm{S}_{\br+\bx}
+\bm{S}_{\br}^{T}\cdot \bm{J}_2\cdot\bm{S}_{\br+\by}\Big]
+D\sum_{\br}(S^z_{\br})^2,
\label{eq:H-gauged}
\end{equation}
where \(\hat x=(1,0)\), \(\hat y=(0,1)\), and
\begin{align}
\bm{J}_1=
\begin{pmatrix}
 a+b & \lambda & 0\\
 -\lambda & a-b & 0\\
 0 & 0 & c
\end{pmatrix},\quad
\bm{J}_2=
\begin{pmatrix}
0 & \beta+\gamma & 0\\
\beta-\gamma & 0 & 0\\
0 & 0 & \alpha
\end{pmatrix},\quad
\{a,b,c,\alpha,\beta,\gamma,\lambda,D\}\in\mathbb{R}.
\label{eq:Jy-uniform}
\end{align}
The residual spin-only symmetry group is \(\mathcal{C}_2\), generated by \(R_{[001]}(\pi)\).  The tSSG symmetries are generated by the translations \(\mathbb{T}_x,\mathbb{T}_y\), together with the effective mirror
\(\widetilde{\mathcal{M}}_x=(\mathfrak{Z}_{\mathcal{M}_x}(\vec r)R_{[100]}(\pi)\mid\mid \mathcal{M}_x)\), where
\(\mathfrak{Z}_{\mathcal{M}_x}(\vec r)=(R_{[001]}(\pi))^{r_y}\).  The key nonsymmorphic relation is
\begin{equation}
\widetilde M_x\mathbb{T}_y\widetilde M_x^{-1}\mathbb{T}_y^{-1}=R_{[001]}(\pi)\in\mathcal{S}_0.
\label{eq:tssg-nonsymmorphic-relation}
\end{equation}
This relation is the origin of the momentum-space sewing structure derived below.

\subsection{Flavor-wave expansion about the large-\(D\) ferroquadrupolar vacuum}

\subsubsection*{1. Cartesian spin-1 basis and flavor bosons}
We use the Cartesian spin-1 basis
\begin{equation}
|x\rangle=\frac{\ii}{\sqrt2}\Big(|1\rangle-|-1\rangle\Big),\qquad
|y\rangle=\frac{1}{\sqrt2}\Big(|1\rangle+|-1\rangle\Big),\qquad
|z\rangle=-\ii|0\rangle.
\label{eq:cartesian-basis}
\end{equation}
In this basis, the single-ion anisotropy is
\((S^z)^2=|x\rangle\langle x|+|y\rangle\langle y|\).  Hence, for large positive \(D\), the unique local minimum is \(|z\rangle\), and the corresponding ferroquadrupolar product state is
\begin{equation}
|\mathrm{FQ}\rangle=\prod_{\br}|z\rangle_{\br}.
\label{eq:FQ-vacuum}
\end{equation}
To expand about this state, introduce three Schwinger bosons \(b_{x,\br},b_{y,\br},b_{z,\br}\) subject to the single-occupancy constraint
\begin{equation}
 b_{x,\br}^{\dagger}b_{x,\br}+b_{y,\br}^{\dagger}b_{y,\br}+b_{z,\br}^{\dagger}b_{z,\br}=1.
\label{eq:constraint}
\end{equation}
The spin operators in the Cartesian basis are
\begin{equation}
S^{m_1}_{\br}=-\ii\,\epsilon_{m_1m_2m_3}
 b^{\dagger}_{m_2,\br}b_{m_3,\br},
\qquad m_1,m_2,m_3\in\{x,y,z\}.
\label{eq:spin-cartesian-bosons}
\end{equation}
Expanding around the large-\(D\) vacuum amounts to condensing the \(z\) flavor,
\begin{equation}
 b_{z,\br}=\sqrt{1-n_{x,\br}-n_{y,\br}}
 \simeq 1-\frac12\big(n_{x,\br}+n_{y,\br}\big),
\label{eq:bz-condensate}
\end{equation}
where \(n_{m_1,\br}=b^{\dagger}_{m_1,\br}b_{m_1,\br}\).  To harmonic order, the spin operators reduce to
\begin{align}
&S^x_{\br}=-\ii\big(b^{\dagger}_{y,\br}-b_{y,\br}\big)+\mathcal{O}(b^3),
\label{eq:Sx-linear}
\\
&S^y_{\br}=\ii\big(b^{\dagger}_{x,\br}-b_{x,\br}\big)+\mathcal{O}(b^3),
\label{eq:Sy-linear}
\\
&S^z_{\br}=-\ii\big(b^{\dagger}_{x,\br}b_{y,\br}-b^{\dagger}_{y,\br}b_{x,\br}\big)
=\mathcal{O}(b^2),
\label{eq:Sz-quadratic}
\\
&(S^z_{\br})^2=n_{x,\br}+n_{y,\br}.
\label{eq:Sz2-exact-boson}
\end{align}
Equation~\eqref{eq:Sz2-exact-boson} has an immediate consequence: all exchange terms involving \(S^z\) are at least quartic in the transverse bosons.  Therefore the couplings \(c\) and \(\alpha\) drop out of the harmonic Hamiltonian.

The local quadrupoles in the \(\{Q_{xz},Q_{yz}\}\) sector are
\begin{equation}
Q_{xz}=-(|x\rangle\langle z|+|z\rangle\langle x|),
\qquad
Q_{yz}=-(|y\rangle\langle z|+|z\rangle\langle y|).
\label{eq:Qxz-Qyz-projectors}
\end{equation}
Using Eq.~\eqref{eq:bz-condensate}, we obtain
\begin{equation}
Q_{xz,\br}= -\big(b_{x,\br}+b^{\dagger}_{x,\br}\big)+\mathcal{O}(b^3),
\qquad
Q_{yz,\br}= -\big(b_{y,\br}+b^{\dagger}_{y,\br}\big)+\mathcal{O}(b^3).
\label{eq:Q-linear-first-group}
\end{equation}
Thus the \(\{Q_{xz},Q_{yz}\}\) sector is already a genuine one-boson sector at harmonic order.

\subsubsection*{2. Quadratic form in real space}

Introduce the canonical quadratures
\begin{equation}
X_{m_1,\br}=\frac{b_{m_1,\br}+b_{m_1,\br}^{\dagger}}{\sqrt2},
\qquad
P_{m_1,\br}=\frac{b_{m_1,\br}-b_{m_1,\br}^{\dagger}}{\ii\sqrt2},
\qquad m_1=x,y.
\label{eq:quadratures}
\end{equation}
Equations~\eqref{eq:Sx-linear} and \eqref{eq:Sy-linear} then become
\begin{equation}
S^x_{\br}=-\sqrt2\,P_{y,\br},
\qquad
S^y_{\br}=+\sqrt2\,P_{x,\br}.
\label{eq:S-in-P}
\end{equation}
The single-ion term is
\begin{equation}
H_{\text{ion}}=D\sum_{\br}(S^z_{\br})^2
=D\sum_{\br}(n_{x,\br}+n_{y,\br})
=\frac{D}{2}\sum_{\br}\Big(X_{\br}^{T}X_{\br}+P_{\br}^{T}P_{\br}\Big)+\mathrm{const.},
\label{eq:D-term-quadratures}
\end{equation}
where \(X_{\br}=(X_{x,\br},X_{y,\br})^T\) and \(P_{\br}=(P_{x,\br},P_{y,\br})^T\).  Substituting Eq.~\eqref{eq:S-in-P} into the horizontal bond gives
\begin{equation}
H_h=\bm{S}_{\br}^{T}\bm{J}_1\bm{S}_{\br+\bx}-cS_{\br}^zS_{\br+\bx}^z
=P_{\br}^{T} C_x P_{\br+\bx},
\qquad
C_x=2
\begin{pmatrix}
 a-b & \lambda\\
 -\lambda & a+b
\end{pmatrix}.
\label{eq:Cx-def}
\end{equation}
Similarly, the vertical bond gives
\begin{equation}
H_v=\bm{S}_{\br}^{T}\bm{J}_2\bm{S}_{\br+\by}-\alpha S_{\br}^zS_{\br+\by}^z
=P_{\br}^{T} C_y P_{\br+\by},
\qquad
C_y=
\begin{pmatrix}
0 & -2\beta+2\gamma\\
-2\beta-2\gamma & 0
\end{pmatrix}.
\label{eq:Cy-def}
\end{equation}
Collecting all harmonic terms, the \(\{Q_{xz},Q_{yz}\}\)-sector Hamiltonian is
\begin{equation}
H_1=H_{\text{ion}}+H_h+H_v
=\frac12\sum_{\br}\Big[D\,X_{\br}^{T}X_{\br}+D\,P_{\br}^{T}P_{\br}\Big]
+\sum_{\br}\Big[P_{\br}^{T}C_xP_{\br+\bx}+P_{\br}^{T}C_yP_{\br+\by}\Big].
\label{eq:H1-realspace}
\end{equation}
This real-space form already exposes the main simplification: the exchange couplings enter only through the \(P\) sector, while the \(X\) sector has the trivial stiffness \(D\mathbb{I}_2\).

\subsubsection*{3. Momentum-space canonical bosonic Hamiltonian}

With the Fourier convention
\begin{equation}
X_{\br}=\frac{1}{\sqrt N}\sum_{\bk}\ee^{\ii\bk\cdot\br}X_{\bk},
\qquad
P_{\br}=\frac{1}{\sqrt N}\sum_{\bk}\ee^{\ii\bk\cdot\br}P_{\bk},
\qquad
\bk=(k_x,k_y)\in\mathrm{BZ},
\label{eq:FT-quadratures}
\end{equation}
Eq.~\eqref{eq:H1-realspace} becomes
\begin{equation}
H_1=\frac12\sum_{\bk}\Big[X_{-\bk}^{\dagger}(D\mathbb{I}_2)X_{\bk}
+P_{-\bk}^{\dagger}M_1(\bk)P_{\bk}\Big],
\label{eq:H1-canonical-kspace}
\end{equation}
with
\begin{align}
M_1(\bk)=D\mathbb{I}_2+4a\cos k_x\,\mathbb{I}_2
-4\beta\cos k_y\,\sigma_x
+4\big(\lambda\sin k_x+\gamma\sin k_y\big)\sigma_y
-4b\cos k_x\,\sigma_z.
\label{eq:M1-def}
\end{align}
This \(2\times2\) Hermitian matrix is the central object of the harmonic theory in the \(\{Q_{xz},Q_{yz}\}\) sector.  Since the \(X\) block is simply \(D\mathbb{I}_2\), the normal-mode problem is completely determined by the Hermitian matrix \(M_1(\bk)\).

For completeness, the same Hamiltonian can also be written in the standard Nambu basis
\begin{equation}
\Psi_{\bk}=
\begin{pmatrix}
 b_{x,\bk}, b_{y,\bk}, b^{\dagger}_{x,-\bk}, b^{\dagger}_{y,-\bk}
\end{pmatrix}^T,
\qquad
H_1=\frac12\sum_{\bk}\Psi_{\bk}^{\dagger}\,\cH_{\mathrm{BdG}}(\bk)\,\Psi_{\bk}+\mathrm{const.},
\label{eq:Nambu-basis}
\end{equation}
where
\begin{equation}
\cH_{\mathrm{BdG}}(\bk)=
\frac{1}{2}\begin{pmatrix}
D\mathbb{I}_2+M_1(\bk) & D\mathbb{I}_2-M_1(\bk)\\
D\mathbb{I}_2-M_1(\bk) & D\mathbb{I}_2+M_1(\bk)
\end{pmatrix}.
\label{eq:Hbdg-block}
\end{equation}
Thus the Nambu representation in Eq.~\eqref{eq:Hbdg-block} is algebraically equivalent to the canonical quadrature representation in Eq.~\eqref{eq:H1-canonical-kspace}.  We therefore diagonalize the problem through \(M_1(\bk)\), where the structure is most transparent.

\subsection{Analytic eigenvalue solutions of the bulk bosonic Hamiltonian}

Define
\begin{equation}
\xi(\bk)=\lambda\sin k_x+\gamma\sin k_y,
\qquad
\rho(\bk)=\sqrt{b^2\cos^2 k_x+\beta^2\cos^2 k_y+\xi(\bk)^2}.
\label{eq:rho-def}
\end{equation}
Then Eq.~\eqref{eq:M1-def} can be written as
\begin{equation}
M_1(\bk)=m_0(\bk)\,\mathbb{I}_2+\bm{d}(\bk)\cdot\bm{\sigma},
\label{eq:M1-dvector}
\end{equation}
with
\begin{equation}
 m_0(\bk)=D+4a\cos k_x,
\qquad
\bm{d}(\bk)=\big(-4\beta\cos k_y,\,4\xi(\bk),\,-4b\cos k_x\big),
\qquad
|\bm{d}(\bk)|=4\rho(\bk).
\label{eq:dvector-components}
\end{equation}
The two eigenvalues of \(M_1(\bk)\) are therefore
\begin{equation}
\mu_{\pm}(\bk)=D+4a\cos k_x\pm 4\rho(\bk).
\label{eq:mu-pm}
\end{equation}
In the stable large-\(D\) regime, each normal mode is a harmonic oscillator with frequency
\begin{equation}
\omega_{\pm}(\bk)=\sqrt{D\,\mu_{\pm}(\bk)}
=\sqrt{D\Big[D+4a\cos k_x\pm 4\rho(\bk)\Big]}.
\label{eq:omega-pm}
\end{equation}
These are the two physical one-particle branches in the \(\{Q_{xz},Q_{yz}\}\) sector.  In the bosonic BdG language, the spectrum contains the particle-hole-related quartet
\(\{+\omega_{-}(\bk),+\omega_{+}(\bk),-\omega_{-}(\bk),-\omega_{+}(\bk)\}\), but only the positive pair corresponds to independent physical excitations.

\subsection{Detailed calculation of the glide-reflection sewing structure}

Recall that
\(\widetilde{\mathcal{M}}_x=(\mathfrak{Z}_{\mathcal{M}_x}(\vec r)R_x(\pi)\mid\mid \mathcal{M}_x)\), with
\(\mathfrak{Z}_{\mathcal{M}_x}(\vec r)=(R_{[001]}(\pi))^{r_y}\).  Denote \(U_M=-\sigma_z\).  The actions of \(\mathbb{T}_y\) and \(\widetilde{\mathcal{M}}_x\) on the boson doublet
\(\bm{b}_{\vec r}=(b_{x,\vec r},b_{y,\vec r})^T\) are
\begin{align}
\mathbb{T}_y:\quad &\bm{b}_{(r_x,r_y)}\mapsto \bm{b}_{(r_x,r_y+1)},
\\
\Mx:\quad &\bm{b}_{(r_x,r_y)}
\mapsto (-1)^{r_y}U_M\bm{b}_{(1-r_x,r_y)}.
\label{eq:Mx-realspace-gauged}
\end{align}
The extra factor \((-1)^{r_y}\) is exactly what turns the mirror action into a glide-like operation in the Brillouin zone.

Using the Fourier convention
\begin{equation}
\bm{b}_{\bk}=\frac{1}{\sqrt N}\sum_{r_x,r_y}\ee^{-\ii(k_xr_x+k_yr_y)}\bm{b}_{(r_x,r_y)},
\label{eq:FT-boson}
\end{equation}
Eq.~\eqref{eq:Mx-realspace-gauged} gives
\begin{equation}
\Mx:\quad \bm{b}_{(k_x,k_y)}
\mapsto \ee^{-\ii k_x}U_M\bm{b}_{(-k_x,\,k_y+\pi)}.
\label{eq:Mx-kspace-action}
\end{equation}
The phase \(\ee^{-\ii k_x}\) is a geometric phase due solely to the mirror axis at \(x=\frac12+n\), \(n\in\mathbb{Z}\).  The physically nontrivial part is the momentum shift
\begin{equation}
(k_x,k_y)\mapsto (-k_x,k_y+\pi).
\label{eq:glide-sewing-map}
\end{equation}
Thus the tSSG identification generated by \(\widetilde M_x\) quotients the Brillouin torus into an effective momentum manifold with Klein-bottle topology, rather than leaving an ordinary torus.

The matrix in Eq.~\eqref{eq:M1-def} satisfies the explicit sewing relation
\begin{equation}
M_1(-k_x,k_y+\pi)=U_M\,M_1(k_x,k_y)\,U_M^{-1}.
\label{eq:M1-glide-sewing}
\end{equation}
The proof is direct.  Using
\begin{equation}
\cos(k_y+\pi)=-\cos k_y,
\qquad
\sin(-k_x)=-\sin k_x,
\qquad
\sin(k_y+\pi)=-\sin k_y,
\label{eq:trig-identities}
\end{equation}
we find
\begin{align}
M_1(-k_x,k_y+\pi)
&=D\mathbb{I}_2+4a\cos k_x\,\mathbb{I}_2
+4\beta\cos k_y\,\sigma_x
-4\big(\lambda\sin k_x+\gamma\sin k_y\big)\sigma_y
-4b\cos k_x\,\sigma_z.
\label{eq:M1-shifted}
\end{align}
On the other hand,
\begin{equation}
U_M\sigma_xU_M^{-1}=-\sigma_x,
\qquad
U_M\sigma_yU_M^{-1}=-\sigma_y,
\qquad
U_M\sigma_zU_M^{-1}=+\sigma_z,
\label{eq:UM-Pauli-conjugation}
\end{equation}
and therefore
\begin{align}
U_M M_1(k_x,k_y)U_M^{-1}
&=D\mathbb{I}_2+4a\cos k_x\,\mathbb{I}_2
+4\beta\cos k_y\,\sigma_x
-4\big(\lambda\sin k_x+\gamma\sin k_y\big)\sigma_y
-4b\cos k_x\,\sigma_z.
\label{eq:UMM1UM}
\end{align}
Equations~\eqref{eq:M1-shifted} and \eqref{eq:UMM1UM} coincide, proving Eq.~\eqref{eq:M1-glide-sewing}.  Consequently,
\begin{equation}
\omega_{\nu}(k_x,k_y)=\omega_{\nu}(-k_x,k_y+\pi),
\qquad \nu=\pm.
\label{eq:band-sewing}
\end{equation}

\subsection{Exact ribbon solution for open boundary conditions along \(x\)}

We now fix \(k_y\) and open the system along the \(x\) direction.  The \(P\)-sector eigenproblem following from Eq.~\eqref{eq:H1-realspace} is
\begin{equation}
 t^{\dagger}\psi_{x-1}+h_0(k_y)\psi_x+t\psi_{x+1}=\mu\,\psi_x,
\qquad x=1,2,\dots,
\label{eq:ribbon-general-recursion}
\end{equation}
with the left-edge boundary condition \(\psi_0=0\), where
\begin{equation}
 h_0(k_y)=D\mathbb{I}_2-4\beta\cos k_y\,\sigma_x+4\gamma\sin k_y\,\sigma_y,
\qquad
 t=2a\mathbb{I}_2-2b\sigma_z+2\ii\lambda\sigma_y.
\label{eq:h0-t-general}
\end{equation}
Inserting the Bloch ansatz \(\psi_x=\ee^{\ii k_xx}\phi\) reproduces \(M_1(\bk)\), as required.

A particularly transparent analytic solution appears on the line
\begin{equation}
 a=0,
 \qquad b=\lambda,
\label{eq:solvable-line}
\end{equation}
with \(\lambda\neq0\).  Introduce the unitary rotation
\begin{equation}
W=\frac{1}{\sqrt2}
\begin{pmatrix}
1&1\\
1&-1
\end{pmatrix},
\qquad
\varphi_x=W^{\dagger}\psi_x=
\begin{pmatrix}
A_x\\ B_x
\end{pmatrix}.
\label{eq:W-rotation}
\end{equation}
In this basis,
\begin{align}
 h_0'(k_y)&=W^{\dagger}h_0(k_y)W
 =
\begin{pmatrix}
 D-4\beta\cos k_y & -4\ii\gamma\sin k_y\\
 4\ii\gamma\sin k_y & D+4\beta\cos k_y
\end{pmatrix},
\label{eq:h0-prime}
\\
 t'&=W^{\dagger}tW
 =
\begin{pmatrix}
0 & -4\lambda\\
0 & 0
\end{pmatrix}.
\label{eq:t-prime}
\end{align}
The hopping matrix is strictly triangular.  This triangularity is the algebraic reason the edge problem can be solved exactly.  The real-space recursion becomes
\begin{align}
\big(D-4\beta\cos k_y-\mu\big)A_x-4\ii\gamma\sin k_y\,B_x-4\lambda B_{x+1}&=0,
\label{eq:A-recursion}
\\
4\ii\gamma\sin k_y\,A_x+\big(D+4\beta\cos k_y-\mu\big)B_x-4\lambda A_{x-1}&=0.
\label{eq:B-recursion}
\end{align}

\subsubsection*{1. Semi-infinite left edge}

For the semi-infinite left half-space \(x=1,2,\dots\), impose \(A_0=0\) and seek a normalizable solution.  Equation~\eqref{eq:B-recursion} is solved identically by choosing
\begin{equation}
A_x\equiv 0,
\qquad
\mu_L(k_y)=D+4\beta\cos k_y.
\label{eq:left-mu-choice}
\end{equation}
The remaining equation, Eq.~\eqref{eq:A-recursion}, reduces to the first-order recursion
\begin{equation}
B_{x+1}=r(k_y)\,B_x,
\qquad
r(k_y)=-\ii\,\frac{\gamma\sin k_y}{\lambda}.
\label{eq:left-recursion}
\end{equation}
Thus the left-edge wavefunction is
\begin{equation}
\varphi_x^{L}(k_y)=\mathcal{N}_L(k_y)
\,r(k_y)^{x-1}
\begin{pmatrix}
0\\ 1
\end{pmatrix},
\qquad
\mathcal{N}_L(k_y)=\sqrt{1-|r(k_y)|^2},
\label{eq:left-wavefunction-rotated}
\end{equation}
which is normalizable if and only if
\begin{equation}
|r(k_y)|<1
\quad\Longleftrightarrow\quad
|\gamma\sin k_y|<|\lambda|.
\label{eq:left-normalizability}
\end{equation}
The corresponding physical bosonic energy is
\begin{equation}
\omega_L(k_y)=\sqrt{D\,\mu_L(k_y)}
=\sqrt{D\Big(D+4\beta\cos k_y\Big)}.
\label{eq:omega-left}
\end{equation}

\subsubsection*{2. Semi-infinite right edge}

For a semi-infinite right edge, the calculation is identical after reversing the chain.  Equivalently, on a finite strip of width \(L_x\), one may solve inward from the right boundary.  The exponentially localized right-edge solution is
\begin{equation}
B_x\equiv 0,
\qquad
\mu_R(k_y)=D-4\beta\cos k_y,
\label{eq:right-mu-choice}
\end{equation}
with recursion
\begin{equation}
A_{x-1}=r_R(k_y)\,A_x,
\qquad
r_R(k_y)=\ii\,\frac{\gamma\sin k_y}{\lambda}.
\label{eq:right-recursion}
\end{equation}
Hence
\begin{equation}
\varphi_x^{R}(k_y)=\mathcal{N}_R(k_y)
\,r_R(k_y)^{L_x-x}
\begin{pmatrix}
1\\ 0
\end{pmatrix},
\qquad
\mathcal{N}_R(k_y)=\sqrt{1-|r_R(k_y)|^2},
\label{eq:right-wavefunction-rotated}
\end{equation}
with the same normalizability condition \(|\gamma\sin k_y|<|\lambda|\).  The physical energy is
\begin{equation}
\omega_R(k_y)=\sqrt{D\,\mu_R(k_y)}
=\sqrt{D\Big(D-4\beta\cos k_y\Big)}.
\label{eq:omega-right}
\end{equation}
For a finite strip, the two semi-infinite edge solutions hybridize only through terms of order \(|r(k_y)|^{L_x}\).  Therefore Eqs.~\eqref{eq:omega-left} and \eqref{eq:omega-right} describe the ribbon branches exponentially accurately at large \(L_x\).

\begin{theorem}[Nonlocal momentum twist relation]
The semi-infinite edge solutions derived above obey
\begin{equation}
\mu_L(k_y)=\mu_R(k_y+\pi),
\qquad
r(k_y)=r_R(k_y+\pi),
\label{eq:edge-sewing-mu-r}
\end{equation}
and hence
\begin{equation}
\omega_L(k_y)=\omega_R(k_y+\pi).
\label{eq:edge-Mobius-sewing}
\end{equation}
This is the nonlocal momentum sewing relation of the ribbon spectrum: the left edge at momentum \(k_y\) is sewn to the right edge at momentum \(k_y+\pi\).
\end{theorem}

\subsection{Auxiliary chiral winding and the Klein-bottle $\mathbb{Z}_2$ invariant}

Still on the solvable line \(a=0\), \(b=\lambda\), the rotated Bloch matrix is
\begin{equation}
M_1'(k_x;k_y)=W^{\dagger}M_1(k_x;k_y)W
=
\begin{pmatrix}
 D-4\beta\cos k_y & q(k_x;k_y)\\
 q(k_x;k_y)^{*} & D+4\beta\cos k_y
\end{pmatrix},
\label{eq:Mtilde-chiral-form}
\end{equation}
with
\begin{equation}
W=\frac{1}{\sqrt2}
\begin{pmatrix}
1&1\\
1&-1
\end{pmatrix},
\qquad
q(k_x;k_y)=-4\Big(\lambda\ee^{\ii k_x}
+\ii\gamma\sin k_y\Big).
\label{eq:q-def}
\end{equation}
Separating the diagonal part
\(D\mathbb{I}_2-4\beta\cos k_y\,\tau_z\), the off-diagonal sector has the standard one-dimensional chiral, or AIII/SSH-like, form
\begin{equation}
\cQ(k_x;k_y)=
\begin{pmatrix}
0 & q(k_x;k_y)\\
q(k_x;k_y)^{*} & 0
\end{pmatrix},
\qquad
\tau_z\cQ\tau_z=-\cQ.
\label{eq:Q-chiral}
\end{equation}
The full fixed-\(k_y\) slice \(M_1'\) is not strictly chiral when \(\beta\neq0\), because of the diagonal \(\tau_z\) term. Nevertheless, on the exactly solvable line, the existence and normalizability of the ribbon modes are controlled entirely by the same off-diagonal polynomial \(q\), as we now show.

Let \(z=\ee^{\ii k_x}\). Equation~\eqref{eq:q-def} becomes
\begin{equation}
q(z;k_y)=-4\big(\lambda z+\ii\gamma\sin k_y\big).
\label{eq:q-z}
\end{equation}
When \(q(k_x;k_y)\) is nonzero on the Brillouin-zone circle, the winding number of the auxiliary chiral block is
\begin{equation}
\nu(k_y)=
\frac{1}{2\pi\ii}
\int_{-\pi}^{\pi}\dd k_x\,
\partial_{k_x}\log q(k_x;k_y)
=
\frac{1}{2\pi\ii}
\oint_{|z|=1}\dd z\,
\partial_z\log q(z;k_y).
\label{eq:winding-def}
\end{equation}
By the argument principle, \(\nu(k_y)\) counts the zeros of \(q(z;k_y)\) inside the unit disk. Since the polynomial \(q\) has the single zero
\begin{equation}
z_{*}(k_y)=-\ii\,\frac{\gamma\sin k_y}{\lambda},
\label{eq:z-star}
\end{equation}
one obtains
\begin{equation}
\nu(k_y)=
\begin{cases}
1, & |\gamma\sin k_y|<|\lambda|,\\
0, & |\gamma\sin k_y|>|\lambda|.
\end{cases}
\label{eq:winding-result}
\end{equation}
At \(|\gamma\sin k_y|=|\lambda|\), the zero \(z_*\) lies on the unit circle, the localization length diverges, and the exact edge mode merges with the bulk. Geometrically, as \(k_x\) runs from \(-\pi\) to \(\pi\), \(q(k_x;k_y)\) traces a circle of radius \(4|\lambda|\), centered at \(-4\ii\gamma\sin k_y\). Its winding is \(1\) precisely when the circle encloses the origin, reproducing Eq.~\eqref{eq:winding-result}.

We now connect this auxiliary winding directly to the ribbon states. The real-space left-edge recursion in Eq.~\eqref{eq:left-recursion} is the boundary realization of
\begin{equation}
q(z;k_y)=0.
\label{eq:q-root-equation}
\end{equation}
Indeed, substituting \(B_x\propto z^{x-1}\) and \(A_x=0\) into Eq.~\eqref{eq:A-recursion} gives
\begin{equation}
\big[-4\ii\gamma\sin k_y-4\lambda z\big]B_x=0,
\label{eq:left-root-from-recursion}
\end{equation}
which is precisely Eq.~\eqref{eq:q-root-equation}. Normalizability requires \(|z|<1\); hence \(\nu(k_y)\) equals the number of normalizable left-edge solutions. It is therefore the root-counting invariant of the exactly solvable boundary problem.

Because of the diagonal \(\tau_z\) term, \(\nu(k_y)\) is not a chiral invariant of the full bosonic slice when \(\beta\neq0\). It is instead the invariant of the auxiliary off-diagonal block \(q\). On the solvable line, this distinction does not affect the exact boundary criterion: the \(\beta\) term shifts the edge eigenvalues to
\(\mu_{L,R}(k_y)=D\pm4\beta\cos k_y\), but leaves \(z_*(k_y)\) and the decay factors unchanged. Thus \(\nu(k_y)\) controls exact edge localization, while the glide enforces the nonlocal sewing relation
\begin{equation}
\omega_L(k_y)=\omega_R(k_y+\pi).
\label{eq:edge-sewing-repeat}
\end{equation}

We next relate this slice invariant to the genuine two-dimensional bulk topology. Let
\begin{equation}
M_1'(k_x,k_y)|u_-(k_x,k_y)\rangle
=
\mu_-(k_x,k_y)|u_-(k_x,k_y)\rangle
\label{eq:lower-kernel-band}
\end{equation}
denote the lower eigenline, with positive frequency
\(\omega_-=\sqrt{D\mu_-}\). We assume the dynamically stable and interband-gapped regime
\(\mu_+>\mu_->0\) throughout the Brillouin zone. Since the \(X\)-sector stiffness is \(D\mathbb I_2\), the corresponding positive-frequency normal-mode bundle is isomorphic to the eigenline bundle of \(M_1'\). Its Berry data can therefore be computed from
\begin{equation}
\mathcal A_i
=
\ii\langle u_-|\partial_{k_i}u_-\rangle,
\qquad
\mathcal F_{xy}
=
\partial_{k_x}\mathcal A_y
-
\partial_{k_y}\mathcal A_x,
\qquad
\Gamma(k_y)
=
\int_{-\pi}^{\pi}\dd k_x\,\mathcal A_x.
\label{eq:berry-data-lower-band}
\end{equation}

On the half Brillouin torus
\(\tau_{1/2}=[-\pi,\pi)\times[-\pi,0]\), the Klein-bottle invariant of the lower positive-frequency band is~\cite{chen2022brillouin}
\begin{equation}
\nu_{\mathrm K}
=
\frac{1}{2\pi}
\int_{\tau_{1/2}}\dd^2k\,\mathcal F_{xy}
+
\frac{\Gamma(0)}{\pi}
\pmod 2.
\label{eq:KB-invariant}
\end{equation}
Choosing a smooth gauge that is periodic in \(k_x\) over the cylinder \(\tau_{1/2}\), Stokes' theorem gives
\begin{equation}
\int_{\tau_{1/2}}\dd^2k\,\mathcal F_{xy}
=
\Gamma(-\pi)-\Gamma(0),
\end{equation}
and hence
\begin{equation}
\nu_{\mathrm K}
=
\frac{\Gamma(0)+\Gamma(-\pi)}{2\pi}
\pmod 2,
\label{eq:KB-Zak-flow}
\end{equation}
where \(\Gamma(k_y)\) is followed as a continuous real-valued branch from \(k_y=-\pi\) to \(0\).

The present kernel obeys both time-reversal symmetry and momentum-space glide reflection,
\begin{equation}
M_1'(-k_x,-k_y)
=
M_1'(k_x,k_y)^{*},
\qquad
M_1'(-k_x,k_y+\pi)
=
\tau_xM_1'(k_x,k_y)\tau_x.
\label{eq:TR-glide-kernel}
\end{equation}
The scalar geometric phase in the full glide representation changes the Zak phase only by an integer multiple of \(2\pi\). Equation~\eqref{eq:TR-glide-kernel} therefore implies
\begin{equation}
\Gamma(-k_y)=\Gamma(k_y)\pmod{2\pi},
\qquad
\Gamma(k_y+\pi)=-\Gamma(k_y)
\pmod{2\pi}.
\label{eq:TR-glide-Zak}
\end{equation}
For the continuous branch on \([-\pi,0]\), the combination
\[
\Gamma\left(-\frac{\pi}{2}+\delta\right)
+
\Gamma\left(-\frac{\pi}{2}-\delta\right)
\]
is a continuous integer multiple of \(2\pi\), and is therefore independent of \(\delta\). Evaluating it at \(\delta=0\) and \(\delta=\pi/2\) gives
\begin{equation}
\Gamma(0)+\Gamma(-\pi)
=
2\Gamma\left(-\frac{\pi}{2}\right),
\end{equation}
so that
\begin{equation}
\nu_{\mathrm K}
=
\frac{\Gamma(-\pi/2)}{\pi}
\pmod 2.
\label{eq:KB-central-Zak}
\end{equation}

At \(k_y=-\pi/2\), the diagonal \(\tau_z\) term vanishes and
\begin{equation}
M_1'\left(k_x;-\frac{\pi}{2}\right)
=
D\mathbb I_2
+
\cQ\left(k_x;-\frac{\pi}{2}\right),
\qquad
q\left(k_x;-\frac{\pi}{2}\right)
=
-4\big(\lambda\ee^{\ii k_x}-\ii\gamma\big).
\label{eq:center-chiral-slice}
\end{equation}
Provided that \(q(k_x;-{\pi\over 2})\neq0\) on the Brillouin-zone circle, write
\begin{equation}
q\left(k_x;-\frac{\pi}{2}\right)
=
\left|q\left(k_x;-\frac{\pi}{2}\right)\right|
\ee^{\ii\varphi(k_x)},
\label{eq:center-q-phase}
\end{equation}
where \(\varphi(k_x)\) is chosen as a continuous real-valued lift along
\(k_x\in[-\pi,\pi]\). The two eigenvalues are
\(\mu_{\pm}=D\pm|q|\), and a normalized periodic eigenvector of the lower band can be chosen as
\begin{equation}
|u_-(k_x)\rangle
=
\frac{1}{\sqrt2}
\begin{pmatrix}
-\ee^{\ii\varphi(k_x)}\\
1
\end{pmatrix}.
\label{eq:center-lower-eigenvector}
\end{equation}
Although the continuous phase satisfies \(\varphi(\pi)-\varphi(-\pi)=2\pi\nu\left(-\frac{\pi}{2}\right)\), the eigenvector in Eq.~\eqref{eq:center-lower-eigenvector} is periodic. Its Berry connection is
\begin{equation}
\mathcal A_x(k_x)
=
\ii\langle u_-(k_x)|\partial_{k_x}u_-(k_x)\rangle
=
-\frac{1}{2}\partial_{k_x}\varphi(k_x).
\label{eq:center-berry-connection}
\end{equation}
Consequently, the Zak phase obeys
\begin{equation}
\begin{split}
\Gamma\left(-\frac{\pi}{2}\right)
=
\int_{-\pi}^{\pi}\dd k_x\,\mathcal A_x(k_x)=
-\frac{1}{2}\big[\varphi(\pi)-\varphi(-\pi)\big]=
-\pi\nu\left(-\frac{\pi}{2}\right)
=
\pi\nu\left(-\frac{\pi}{2}\right)
\pmod{2\pi}.
\end{split}
\label{eq:center-Zak-winding}
\end{equation}
Combining Eqs.~\eqref{eq:KB-central-Zak} and
\eqref{eq:center-Zak-winding}, we obtain
\begin{equation}
\boxed{\nu_{\mathrm K}
=
\nu\left(-\frac{\pi}{2}\right)
\pmod 2
=
\begin{cases}
1, & |\gamma|<|\lambda|,\\
0, & |\gamma|>|\lambda|.
\end{cases}}
\label{eq:KB-winding-relation}
\end{equation}
The distinction between the slice winding and the two-dimensional bulk invariant is transparent at the unit-circle crossing. The two eigenvalues of \(M_1'(\bk)\) are separated by
\begin{equation}
\mu_+(\bk)-\mu_-(\bk)
=
8\sqrt{
\beta^2\cos^2 k_y+
\left|\lambda\ee^{\ii k_x}+\ii\gamma\sin k_y\right|^2
}.
\label{eq:interband-splitting}
\end{equation}
Thus, when \(q=0\) on \(|z|=1\), the exact edge root reaches
\(|z_*|=1\) and the corresponding boundary mode delocalizes.
For \(\beta\neq0\) and \(k_y\neq\pm\pi/2\), however, the
bulk splitting remains \(8|\beta\cos k_y|>0\). Such
unit-circle crossings change the root count \(\nu(k_y)\) and
delimit the momentum range of the exact edge branches, but
do not constitute a two-dimensional bulk transition.
Upon tuning \(|\gamma|/|\lambda|\) from below to above unity,
the first crossing occurs at \(|\gamma|=|\lambda|\) on the slice \(k_y=-\pi/2\), where
the diagonal mass also vanishes. The two positive-frequency
bulk bands then touch, allowing
\(\nu_{\mathrm K}=\nu(-\pi/2)\pmod 2\) to change. For
\(|\gamma|>|\lambda|\), the two off-center solutions of
\(|\gamma\sin k_y|=|\lambda|\) merely mark the endpoints of
the remaining exact edge segments.

As a concrete example, set \(\lambda=1\) and \(\gamma=0.9\). Then
\begin{equation}
|\gamma\sin k_y|\le 0.9<1=|\lambda|
\qquad
\text{for all }k_y.
\label{eq:all-slices-nontrivial}
\end{equation}
Thus \(\nu(k_y)=1\) for every \(k_y\): its value at
\(k_y=-\pi/2\) gives \(\nu_{\mathrm K}=1\), while its values over all \(k_y\) ensure exact edge localization throughout the spin Brillouin zone. Together with the glide sewing, this yields a complete nonlocal edge pair rather than  fragmentary edge arcs. 

\newpage
\section{Enumeration of spin-point groups $\mathcal{G}$ satisfying $\mathcal{G}/\mathcal{S}_0=PG\times\mathbb{Z}_2^{\mathcal{T}}$}

\subsection{Algorithmic workflow.}

The classification proceeds as follows:\\

\begin{enumerate}
\item \emph{Point group input.} Construct the point group PG explicitly as a finite subgroup of $O(3)$ from a small generating set.

\item \emph{Choice of spin only group \( \mathcal{S}_0 \).} Fix one of the 11 admissible chiral point groups,
\begin{equation}
\mathcal{S}_0\in \{\mathcal{C}_1, \mathcal{C}_2, \mathcal{C}_3, \mathcal{C}_4, \mathcal{C}_6,\mathcal{D}_2, \mathcal{D}_3, \mathcal{D}_4, \mathcal{D}_6, \mathcal{T},\mathcal{O}\}\subset SO(3).
\end{equation}

\item \emph{Normalizer in \( SO(3) \).} Determine \( \mathfrak{N}_{SO(3)}(\mathcal{S}_0) \) (see Table~\ref{tab:o3-normalizer-so3-centralizer-chiral}). For the discrete nonaxial cases $\mathcal{S}_0\in\{\mathcal{D}_2, \mathcal{D}_3, \mathcal{D}_4, \mathcal{D}_6, \mathcal{T},\mathcal{O}\}$, the normalizer is finite and constructed explicitly. For the axial cases $\mathcal{S}_0\in\{ \mathcal{C}_2, \mathcal{C}_3, \mathcal{C}_4, \mathcal{C}_6\}$, one has \( \mathfrak{N}_{SO(3)}(\mathcal{S}_0)\cong O(2) \), which is continuous but admits a finite representative set sufficient for crystallographic enumeration. For the trivial case \( \mathcal{S}_0=\mathcal{C}_1 \), $\mathfrak{N}_{SO(3)}(\mathcal{C}_1)=SO(3)$, and the continuous target is handled via the list of crystallographic finite rotation groups.

\item \emph{Quotient target (see Table~\ref{tab:o3-normalizer-so3-centralizer-chiral}).} Construct
\begin{equation}
Q_0(\mathcal{S}_0)=\mathfrak{N}_{SO(3)}(\mathcal{S}_0)/\mathcal{S}_0. 
\end{equation}
For continuous normalizers, replace the quotient by a finite representative target capturing all crystallographically admissible images. 

\item \emph{Enumeration of group homomorphisms.} Enumerate all maps
\begin{equation}
\hat{R}:\text{PG}\to Q_0(\mathcal{S}_0)
\end{equation}
by brute force over generator images. Candidates are first filtered by order constraints, $\text{Order}\!\bigl(\hat{R}(l_g)\bigr)\mid \text{Order}(l_g)$, and then extended to the full group by a breadth first search over the multiplication graph. Assignments producing inconsistent values are rejected.

\item \emph{Reduction to conjugation inequivalence.} Quotient the raw set by spin transformations. For finite \( Q_0 \),
\begin{equation}
\hat{R}\sim \hat{R}'\iff \exists q\in Q_0:\quad
\hat{R}'(l_g)=q\,\hat{R}(l_g)\,q^{-1}\quad \forall l_g\in \text{PG}.
\label{eq:conj}
\end{equation}
For continuous targets, reduce by character equivalence of the  orthogonal representations,
\begin{equation}
\chi_{\hat{R}}(l_g)=\Tr\,{\hat{R}}(l_g),
\end{equation}
identifying \( \hat{R} \sim \hat{R}' \) when
\begin{equation}
\chi_{\hat{R}}(l_g)=\chi_{\hat{R}'}(l_g)\qquad \forall l_g\in \text{PG}.
\label{eq:character}
\end{equation}
For the present orthogonal representations, character equivalence coincides with conjugacy.

\item \emph{Adding Time-reversal.} 
The condition \( [\mathcal{T},\text{PG}]=0 \) requires
\begin{equation}
-\hat{R}(\mathcal{T})\in \mathfrak{Z}_{Q_0}\!\bigl(\hat{R}(\text{PG})\bigr),
\qquad \left(-\hat{R}(\mathcal{T})\right)^2=E.
\label{eq:qt}
\end{equation}
This yields two classes:
\begin{align}
\text{Class A:} &\quad -\hat{R}(\mathcal{T})=E,\\
\text{Class B:} &\quad -\hat{R}(\mathcal{T})\neq E,\quad
\left(\hat{R}(\mathcal{T})\right)^2=E,\quad
-\hat{R}(\mathcal{T})\in \mathfrak{Z}_{Q_0}\!\bigl(\hat{R}(\text{PG})\bigr),
\end{align}
where nontrivial involutions are counted up to conjugacy within the same centralizer.

\item \emph{Export and tabulation.} See the following tables. 

\end{enumerate}

\begin{table}[h]
\centering
\caption{Normalizers in $O(3)$ and centralizers in $SO(3)$ for the 11 cPGs, with $\mathbb{Z}_2^{-}=\{E,-E\}$. Coset representatives are written as $\pm R_{\hat n}(\theta)$, with $\hat u(\varphi)=[\cos\varphi,\sin\varphi,0]$.}
\label{tab:o3-normalizer-so3-centralizer-chiral}
\resizebox{\textwidth}{!}{
\begin{tabular}{|c|c|c|c|c|}
\hline
$\text{ cPG }$ &
$\mathfrak{N}_{O(3)}(\text{cPG})$ &
$\mathfrak{Z}_{SO(3)}(\text{cPG})$ &
$\mathfrak{N}_{O(3)}(\text{cPG})/\text{cPG}$ &
Coset representatives in $\mathfrak{N}_{O(3)}(\text{cPG})/\text{cPG}$
\\
\hline
$\mathcal{C}_1$ & $O(3)$ & $SO(3)$ & $O(3)$ &
$\{\pm R_{\hat n}(\theta)\},\ \hat n\in S^2,\ \theta\in[0,\pi]$
\\
\hline
$\mathcal{C}_2$ & $\mathcal{D}_{\infty h}$ & $\mathcal{D}_\infty$ &
$((\mathcal{C}_\infty/\mathcal{C}_2)\rtimes \mathcal{C}_2) \times \mathbb{Z}_2^{-}$ &
\begin{tabular}[c]{@{}l@{}}
$\theta\in[0,\pi):\ \{R_z(\theta),\ R_{\hat u(\theta/2)}(\pi),\ -R_z(\theta),\ -R_{\hat u(\theta/2)}(\pi)\}$
\end{tabular}
\\
\hline
$\mathcal{C}_3$ & $\mathcal{D}_{\infty h}$ & $\mathcal{C}_\infty$ &
$((\mathcal{C}_\infty/\mathcal{C}_3)\rtimes \mathcal{C}_2 ) \times \mathbb{Z}_2^{-}$ &
\begin{tabular}[c]{@{}l@{}}
$\theta\in[0,2\pi/3):\ \{R_z(\theta),\ R_{\hat u(\theta/2)}(\pi),\ -R_z(\theta),\ -R_{\hat u(\theta/2)}(\pi)\}$
\end{tabular}
\\
\hline
$\mathcal{C}_4$ & $\mathcal{D}_{\infty h}$ & $\mathcal{C}_\infty$ &
$((\mathcal{C}_\infty/\mathcal{C}_4)\rtimes \mathcal{C}_2 ) \times \mathbb{Z}_2^{-}$ &
\begin{tabular}[c]{@{}l@{}}
$\theta\in[0,\pi/2):\ \{R_z(\theta),\ R_{\hat u(\theta/2)}(\pi),\ -R_z(\theta),\ -R_{\hat u(\theta/2)}(\pi)\}$
\end{tabular}
\\
\hline
$\mathcal{C}_6$ & $\mathcal{D}_{\infty h}$ & $\mathcal{C}_\infty$ &
$((\mathcal{C}_\infty/\mathcal{C}_6)\rtimes \mathcal{C}_2 ) \times \mathbb{Z}_2^{-}$ &
\begin{tabular}[c]{@{}l@{}}
$\theta\in[0,\pi/3):\ \{R_z(\theta),\ R_{\hat u(\theta/2)}(\pi),\ -R_z(\theta),\ -R_{\hat u(\theta/2)}(\pi)\}$
\end{tabular}
\\
\hline
$\mathcal{D}_2$ & $O_h$ & $\mathcal{D}_2$ &
$\mathcal{D}_3 \times \mathbb{Z}_2^{-}$ &
\begin{tabular}[c]{@{}l@{}}
$\{\pm R_z(0),\ \pm R_{[111]}(2\pi/3),\ \pm R_{[111]}(4\pi/3),$\\
$\ \ \pm R_{[110]}(\pi),\ \pm R_{[101]}(\pi),\ \pm R_{[011]}(\pi)\}$
\end{tabular}
\\
\hline
$\mathcal{D}_3$ & $\mathcal{D}_{6h}$ & $\mathcal{C}_2$ &
$\mathcal{C}_2\times \mathbb{Z}_2^{-}$ &
$\{R_z(0),\ R_z(\pi/3),\ -R_z(0),\ -R_z(\pi/3)\}$
\\
\hline
$\mathcal{D}_4$ & $\mathcal{D}_{8h}$ & $\mathcal{C}_2$ &
$\mathcal{C}_2\times  \mathbb{Z}_2^{-}$ &
$\{R_z(0),\ R_z(\pi/4),\ -R_z(0),\ -R_z(\pi/4)\}$
\\
\hline
$\mathcal{D}_6$ & $\mathcal{D}_{12h}$ & $\mathcal{C}_2$ &
$\mathcal{C}_2 \times \mathbb{Z}_2^{-}$ &
$\{R_z(0),\ R_z(\pi/6),\ -R_z(0),\ -R_z(\pi/6)\}$
\\
\hline
$T$ & $O_h$ & $\mathcal{C}_1$ &
$\mathcal{C}_2 \times \mathbb{Z}_2^{-}$ &
$\{R_z(0),\ R_z(\pi/2),\ -R_z(0),\ -R_z(\pi/2)\}$
\\
\hline
$O$ & $O_h$ & $\mathcal{C}_1$ &
$ \mathbb{Z}_2^{-}$ &
$\{R_z(0),\ -R_z(0)\}$
\\
\hline
\end{tabular}}
\end{table}

\begin{table}[h]
\centering
\caption{Counting of physically inequivalent homomorphisms
$\hat{R}:PG\times \mathbb{Z}_2^{\mathcal{T}}\to \mathfrak{N}_{O(3)}(S_0)/S_0$
with $\det\hat{R}(PG)=+1$ and $\det\hat{R}(\mathcal{T})=-1$.
Each entry is $T+nT$ with $T$ denoting the existence of exact time-reversal symmetry ($\hat{R}(\mathcal{T})=-I_3$), and $nT$ denoting the existence of effective time-reversal symmetry ($\hat{R}(\mathcal{T})\neq -I_3$), respectively.}
\label{SM:Table-counting-summary}
\resizebox{\textwidth}{!}{%
\begin{tabular}{|c|c|c|c|c|c|c|c|c|}
\hline
$S_0$ & Triclinic (2) & Monoclinic (3) & Orthorhombic (3) & Tetragonal (7) & Trigonal (5) & Hexagonal (7) & Cubic (5) & Total (32) \\ \hline
$\mathcal{C}_1$ & $3+4$ & $9+16$ & $25+56$ & $48+91$ & $19+27$ & $57+100$ & $24+27$ & $185+321$ \\ \hline
$\mathcal{C}_2$ & $4+9$ & $16+40$ & $56+152$ & $91+227$ & $27+56$ & $100+236$ & $27+56$ & $321+776$ \\ \hline
$\mathcal{C}_3$ & $4+9$ & $16+40$ & $56+152$ & $91+227$ & $27+56$ & $100+236$ & $27+56$ & $321+776$ \\ \hline
$\mathcal{C}_4$ & $4+9$ & $16+40$ & $56+152$ & $91+227$ & $27+56$ & $100+236$ & $27+56$ & $321+776$ \\ \hline
$\mathcal{C}_6$ & $4+9$ & $16+40$ & $56+152$ & $91+227$ & $27+56$ & $100+236$ & $27+56$ & $321+776$ \\ \hline
$\mathcal{D}_2$ & $3+3$ & $8+8$ & $16+16$ & $28+28$ & $16+11$ & $35+28$ & $16+11$ & $122+105$ \\ \hline
$\mathcal{D}_3$ & $3+3$ & $8+8$ & $16+16$ & $28+28$ & $11+11$ & $28+28$ & $11+11$ & $105+105$ \\ \hline
$\mathcal{D}_4$ & $3+3$ & $8+8$ & $16+16$ & $28+28$ & $11+11$ & $28+28$ & $11+11$ & $105+105$ \\ \hline
$\mathcal{D}_6$ & $3+3$ & $8+8$ & $16+16$ & $28+28$ & $11+11$ & $28+28$ & $11+11$ & $105+105$ \\ \hline
$\mathcal{T}$ & $3+3$ & $8+8$ & $16+16$ & $28+28$ & $11+11$ & $28+28$ & $11+11$ & $105+105$ \\ \hline
$\mathcal{O}$ & $2+0$ & $3+0$ & $3+0$ & $7+0$ & $5+0$ & $7+0$ & $5+0$ & $32+0$ \\ \hline
\textbf{Total} & \textbf{36+55} & \textbf{116+216} & \textbf{332+744} & \textbf{559+1139} & \textbf{192+306} & \textbf{611+1184} & \textbf{197+306} & \textbf{2043+3950} \\ \hline
\end{tabular}}
\end{table}

\newpage
\setlength{\LTleft}{-0.5in}
\setlength{\LTright}{-0.5in}
\setlength{\PGTwidth}{\dimexpr\textwidth+1in-8\tabcolsep-5\arrayrulewidth\relax}

\Needspace{8\baselineskip}
\subsection{$S_0=\mathcal{C}_1$}
\subsubsection{1. $S_0=\mathcal{C}_1$: SPGs with exact time-reversal symmetry}
{\scriptsize
\setlength{\LTpre}{0pt}
\setlength{\LTpost}{0pt}
\begin{longtable}{|L{0.10\PGTwidth}|L{0.18\PGTwidth}|L{0.64\PGTwidth}|C{0.08\PGTwidth}|}
\hline
\rowcolor{gray!12}\textbf{PG} & \textbf{PG generators} & \textbf{Representatives (on generators)} & \textbf{\# SPGs}\\
\cline{1-3}
\endfirsthead
\hline
\rowcolor{gray!12}\textbf{PG} & \textbf{PG generators} & \textbf{Representatives (on generators)} & \textbf{\# SPGs}\\
\cline{1-3}
\endhead
\cline{1-3}
\endfoot
\hline
\endlastfoot
$\mathcal{C}_1$ & $\varnothing$ & $\hat{R}_{1}:$ \ (trivial),\allowbreak\ $\bigl(-E\Vert T\bigr)$ & 1 \\ \hline
$\mathcal{C}_i$ & $\mathcal{I}$ & $\hat{R}_{1}:$ \ $\bigl(R_{[001]}\!\left(\pi\right)\Vert \mathcal{I}\bigr)$,\allowbreak\ $\bigl(-E\Vert T\bigr)$ & 2 \\ \cline{1-3}
$\mathcal{C}_i$ & $\mathcal{I}$ & $\hat{R}_{2}:$ \ $\bigl(E\Vert \mathcal{I}\bigr)$,\allowbreak\ $\bigl(-E\Vert T\bigr)$ &  \\ \hline
$\mathcal{C}_2$ & $C_{2z}$ & $\hat{R}_{1}:$ \ $\bigl(R_{[001]}\!\left(\pi\right)\Vert C_{2z}\bigr)$,\allowbreak\ $\bigl(-E\Vert T\bigr)$ & 2 \\ \cline{1-3}
$\mathcal{C}_2$ & $C_{2z}$ & $\hat{R}_{2}:$ \ $\bigl(E\Vert C_{2z}\bigr)$,\allowbreak\ $\bigl(-E\Vert T\bigr)$ &  \\ \hline
$\mathcal{C}_s$ & $\mathcal{M}_x$ & $\hat{R}_{1}:$ \ $\bigl(R_{[001]}\!\left(\pi\right)\Vert \mathcal{M}_x\bigr)$,\allowbreak\ $\bigl(-E\Vert T\bigr)$ & 2 \\ \cline{1-3}
$\mathcal{C}_s$ & $\mathcal{M}_x$ & $\hat{R}_{2}:$ \ $\bigl(E\Vert \mathcal{M}_x\bigr)$,\allowbreak\ $\bigl(-E\Vert T\bigr)$ &  \\ \hline
$\mathcal{C}_{2h}$ & $C_{2z},\ \mathcal{I}$ & $\hat{R}_{1}:$ \ $\bigl(R_{[001]}\!\left(\pi\right)\Vert C_{2z}\bigr)$,\allowbreak\ $\bigl(R_{[100]}\!\left(\pi\right)\Vert \mathcal{I}\bigr)$,\allowbreak\ $\bigl(-E\Vert T\bigr)$ &  \\
\hline
\end{longtable}
}

\newpage
\subsubsection{2. $S_0=\mathcal{C}_1$: SPGs with effective time-reversal symmetry}
{\scriptsize
\setlength{\LTpre}{0pt}
\setlength{\LTpost}{0pt}
\begin{longtable}{|L{0.10\PGTwidth}|L{0.18\PGTwidth}|L{0.64\PGTwidth}|C{0.08\PGTwidth}|}
\hline
\rowcolor{gray!12}\textbf{PG} & \textbf{PG generators} & \textbf{Representatives (on generators)} & \textbf{\# SPGs}\\
\cline{1-3}
\endfirsthead
\hline
\rowcolor{gray!12}\textbf{PG} & \textbf{PG generators} & \textbf{Representatives (on generators)} & \textbf{\# SPGs}\\
\cline{1-3}
\endhead
\cline{1-3}
\endfoot
\hline
\endlastfoot
$\mathcal{C}_1$ & $\varnothing$ & $\hat{R}_{1}:$ \ $\bigl(-R_{[001]}\!\left(\pi\right)\Vert T\bigr)$ & 1 \\ \hline
$\mathcal{C}_i$ & $\mathcal{I}$ & $\hat{R}_{1}:$ \ $\bigl(E\Vert \mathcal{I}\bigr)$,\allowbreak\ $\bigl(-R_{[001]}\!\left(\pi\right)\Vert T\bigr)$ &  \\ \cline{1-3}
$\mathcal{C}_i$ & $\mathcal{I}$ & $\hat{R}_{2}:$ \ $\bigl(R_{[001]}\!\left(\pi\right)\Vert \mathcal{I}\bigr)$,\allowbreak\ $\bigl(-R_{[001]}\!\left(\pi\right)\Vert T\bigr)$ & 3 \\ \cline{1-3}
$\mathcal{C}_i$ & $\mathcal{I}$ & $\hat{R}_{3}:$ \ $\bigl(R_{[100]}\!\left(\pi\right)\Vert \mathcal{I}\bigr)$,\allowbreak\ $\bigl(-R_{[001]}\!\left(\pi\right)\Vert T\bigr)$ &  \\ \hline
$\mathcal{C}_2$ & $C_{2z}$ & $\hat{R}_{1}:$ \ $\bigl(E\Vert C_{2z}\bigr)$,\allowbreak\ $\bigl(-R_{[001]}\!\left(\pi\right)\Vert T\bigr)$ &  \\ \cline{1-3}
$\mathcal{C}_2$ & $C_{2z}$ & $\hat{R}_{2}:$ \ $\bigl(R_{[001]}\!\left(\pi\right)\Vert C_{2z}\bigr)$,\allowbreak\ $\bigl(-R_{[001]}\!\left(\pi\right)\Vert T\bigr)$ & 3 \\ \cline{1-3}
$\mathcal{C}_2$ & $C_{2z}$ & $\hat{R}_{3}:$ \ $\bigl(R_{[100]}\!\left(\pi\right)\Vert C_{2z}\bigr)$,\allowbreak\ $\bigl(-R_{[001]}\!\left(\pi\right)\Vert T\bigr)$ &  \\ \hline
$\mathcal{C}_s$ & $\mathcal{M}_x$ & $\hat{R}_{1}:$ \ $\bigl(E\Vert \mathcal{M}_x\bigr)$,\allowbreak\ $\bigl(-R_{[001]}\!\left(\pi\right)\Vert T\bigr)$ &  \\
\hline
\end{longtable}
}

\newpage

\Needspace{8\baselineskip}
\subsection{$S_0=\mathcal{C}_n$ ($n=2,3,4,6$)}

In this section, spin-only group $S_0$ is generated by the rotation $R_{z}(2\pi/n)$ with $n=2,3,4,6$. 

\subsubsection{1. $S_0=\mathcal{C}_n$: SPGs with exact time-reversal symmetry }
{\scriptsize
\setlength{\LTpre}{0pt}
\setlength{\LTpost}{0pt}
\begin{longtable}{|L{0.10\PGTwidth}|L{0.18\PGTwidth}|L{0.64\PGTwidth}|C{0.08\PGTwidth}|}
\hline
\rowcolor{gray!12}\textbf{PG} & \textbf{PG generators} & \textbf{Representatives (on generators)} & \textbf{\# SPGs}\\
\cline{1-3}
\endfirsthead
\hline
\rowcolor{gray!12}\textbf{PG} & \textbf{PG generators} & \textbf{Representatives (on generators)} & \textbf{\# SPGs}\\
\cline{1-3}
\endhead
\cline{1-3}
\endfoot
\hline
\endlastfoot
$\mathcal{C}_1$ & $\varnothing$ & $\hat{R}_{1}:$ \ (trivial),\allowbreak\ $\bigl(-E\Vert T\bigr)$ & 1 \\ \hline
$\mathcal{C}_i$ & $\mathcal{I}$ & $\hat{R}_{1}:$ \ $\bigl(E\Vert \mathcal{I}\bigr)$,\allowbreak\ $\bigl(-E\Vert T\bigr)$ &  \\ \cline{1-3}
$\mathcal{C}_i$ & $\mathcal{I}$ & $\hat{R}_{2}:$ \ $\bigl(R_{[001]}\!\left(\frac{\pi}{n}\right)\Vert \mathcal{I}\bigr)$,\allowbreak\ $\bigl(-E\Vert T\bigr)$ & 3 \\ \cline{1-3}
$\mathcal{C}_i$ & $\mathcal{I}$ & $\hat{R}_{3}:$ \ $\bigl(R_{[100]}\!\left(\pi\right)\Vert \mathcal{I}\bigr)$,\allowbreak\ $\bigl(-E\Vert T\bigr)$ &  \\ \hline
$\mathcal{C}_2$ & $C_{2z}$ & $\hat{R}_{1}:$ \ $\bigl(E\Vert C_{2z}\bigr)$,\allowbreak\ $\bigl(-E\Vert T\bigr)$ &  \\ \cline{1-3}
$\mathcal{C}_2$ & $C_{2z}$ & $\hat{R}_{2}:$ \ $\bigl(R_{[001]}\!\left(\frac{\pi}{n}\right)\Vert C_{2z}\bigr)$,\allowbreak\ $\bigl(-E\Vert T\bigr)$ & 3 \\ \cline{1-3}
$\mathcal{C}_2$ & $C_{2z}$ & $\hat{R}_{3}:$ \ $\bigl(R_{[100]}\!\left(\pi\right)\Vert C_{2z}\bigr)$,\allowbreak\ $\bigl(-E\Vert T\bigr)$ &  \\ \hline
$\mathcal{C}_s$ & $\mathcal{M}_x$ & $\hat{R}_{1}:$ \ $\bigl(E\Vert \mathcal{M}_x\bigr)$,\allowbreak\ $\bigl(-E\Vert T\bigr)$ &  \\
\hline
\end{longtable}
}

\newpage
\subsubsection{2. $S_0=\mathcal{C}_n$: SPGs with effective time-reversal symmetry }
{\scriptsize
\setlength{\LTpre}{0pt}
\setlength{\LTpost}{0pt}
\begin{longtable}{|L{0.10\PGTwidth}|L{0.18\PGTwidth}|L{0.64\PGTwidth}|C{0.08\PGTwidth}|}
\hline
\rowcolor{gray!12}\textbf{PG} & \textbf{PG generators} & \textbf{Representatives (on generators)} & \textbf{\# SPGs}\\
\cline{1-3}
\endfirsthead
\hline
\rowcolor{gray!12}\textbf{PG} & \textbf{PG generators} & \textbf{Representatives (on generators)} & \textbf{\# SPGs}\\
\cline{1-3}
\endhead
\cline{1-3}
\endfoot
\hline
\endlastfoot
$\mathcal{C}_1$ & $\varnothing$ & $\hat{R}_{1}:$ \ $\bigl(-R_{[001]}\!\left(\frac{\pi}{n}\right)\Vert T\bigr)$ & 2 \\ \cline{1-3}
$\mathcal{C}_1$ & $\varnothing$ & $\hat{R}_{2}:$ \ $\bigl(-R_{[100]}\!\left(\pi\right)\Vert T\bigr)$ &  \\ \hline
$\mathcal{C}_i$ & $\mathcal{I}$ & $\hat{R}_{1}:$ \ $\bigl(E\Vert \mathcal{I}\bigr)$,\allowbreak\ $\bigl(-R_{[001]}\!\left(\frac{\pi}{n}\right)\Vert T\bigr)$ &  \\ \cline{1-3}
$\mathcal{C}_i$ & $\mathcal{I}$ & $\hat{R}_{2}:$ \ $\bigl(E\Vert \mathcal{I}\bigr)$,\allowbreak\ $\bigl(-R_{[100]}\!\left(\pi\right)\Vert T\bigr)$ &  \\ \cline{1-3}
$\mathcal{C}_i$ & $\mathcal{I}$ & $\hat{R}_{3}:$ \ $\bigl(R_{[001]}\!\left(\frac{\pi}{n}\right)\Vert \mathcal{I}\bigr)$,\allowbreak\ $\bigl(-R_{[001]}\!\left(\frac{\pi}{n}\right)\Vert T\bigr)$ &  \\ \cline{1-3}
$\mathcal{C}_i$ & $\mathcal{I}$ & $\hat{R}_{4}:$ \ $\bigl(R_{[001]}\!\left(\frac{\pi}{n}\right)\Vert \mathcal{I}\bigr)$,\allowbreak\ $\bigl(-R_{[100]}\!\left(\pi\right)\Vert T\bigr)$ & 7 \\ \cline{1-3}
$\mathcal{C}_i$ & $\mathcal{I}$ & $\hat{R}_{5}:$ \ $\bigl(R_{[100]}\!\left(\pi\right)\Vert \mathcal{I}\bigr)$,\allowbreak\ $\bigl(-R_{[001]}\!\left(\frac{\pi}{n}\right)\Vert T\bigr)$ &  \\ \cline{1-3}
$\mathcal{C}_i$ & $\mathcal{I}$ & $\hat{R}_{6}:$ \ $\bigl(R_{[100]}\!\left(\pi\right)\Vert \mathcal{I}\bigr)$,\allowbreak\ $\bigl(-R_{[100]}\!\left(\pi\right)\Vert T\bigr)$ &  \\
\hline
\end{longtable}
}

\newpage

\Needspace{8\baselineskip}
\subsection{$S_0=\mathcal{D}_2$}

\subsubsection{1. $S_0=\mathcal{D}_2$: SPGs with exact time-reversal symmetry }

{\scriptsize
\setlength{\LTpre}{0pt}
\setlength{\LTpost}{0pt}
\begin{longtable}{|L{0.10\PGTwidth}|L{0.18\PGTwidth}|L{0.64\PGTwidth}|C{0.08\PGTwidth}|}
\hline
\rowcolor{gray!12}\textbf{PG} & \textbf{PG generators} & \textbf{Representatives (on generators)} & \textbf{\# SPGs}\\
\cline{1-3}
\endfirsthead
\hline
\rowcolor{gray!12}\textbf{PG} & \textbf{PG generators} & \textbf{Representatives (on generators)} & \textbf{\# SPGs}\\
\cline{1-3}
\endhead
\cline{1-3}
\endfoot
\hline
\endlastfoot
$\mathcal{C}_1$ & $\varnothing$ & $\hat{R}_{1}:$ \ (trivial),\allowbreak\ $\bigl(-E\Vert T\bigr)$ & 1 \\ \hline
$\mathcal{C}_i$ & $\mathcal{I}$ & $\hat{R}_{1}:$ \ $\bigl(E\Vert \mathcal{I}\bigr)$,\allowbreak\ $\bigl(-E\Vert T\bigr)$ & 2 \\ \cline{1-3}
$\mathcal{C}_i$ & $\mathcal{I}$ & $\hat{R}_{2}:$ \ $\bigl(R_{[100]}\!\left(\frac{\pi}{2}\right)\Vert \mathcal{I}\bigr)$,\allowbreak\ $\bigl(-E\Vert T\bigr)$ &  \\ \hline
$\mathcal{C}_2$ & $C_{2z}$ & $\hat{R}_{1}:$ \ $\bigl(E\Vert C_{2z}\bigr)$,\allowbreak\ $\bigl(-E\Vert T\bigr)$ & 2 \\ \cline{1-3}
$\mathcal{C}_2$ & $C_{2z}$ & $\hat{R}_{2}:$ \ $\bigl(R_{[100]}\!\left(\frac{\pi}{2}\right)\Vert C_{2z}\bigr)$,\allowbreak\ $\bigl(-E\Vert T\bigr)$ &  \\ \hline
$\mathcal{C}_s$ & $\mathcal{M}_x$ & $\hat{R}_{1}:$ \ $\bigl(E\Vert \mathcal{M}_x\bigr)$,\allowbreak\ $\bigl(-E\Vert T\bigr)$ & 2 \\ \cline{1-3}
$\mathcal{C}_s$ & $\mathcal{M}_x$ & $\hat{R}_{2}:$ \ $\bigl(R_{[100]}\!\left(\frac{\pi}{2}\right)\Vert \mathcal{M}_x\bigr)$,\allowbreak\ $\bigl(-E\Vert T\bigr)$ &  \\ \hline
$\mathcal{C}_{2h}$ & $C_{2z},\ \mathcal{I}$ & $\hat{R}_{1}:$ \ $\bigl(E\Vert C_{2z}\bigr)$,\allowbreak\ $\bigl(E\Vert \mathcal{I}\bigr)$,\allowbreak\ $\bigl(-E\Vert T\bigr)$ &  \\
\hline
\end{longtable}
}

\newpage
\subsubsection{2. $S_0=\mathcal{D}_2$: SPGs with effective time-reversal symmetry }

{\scriptsize
\setlength{\LTpre}{0pt}
\setlength{\LTpost}{0pt}
\begin{longtable}{|L{0.10\PGTwidth}|L{0.18\PGTwidth}|L{0.64\PGTwidth}|C{0.08\PGTwidth}|}
\hline
\rowcolor{gray!12}\textbf{PG} & \textbf{PG generators} & \textbf{Representatives (on generators)} & \textbf{\# SPGs}\\
\cline{1-3}
\endfirsthead
\hline
\rowcolor{gray!12}\textbf{PG} & \textbf{PG generators} & \textbf{Representatives (on generators)} & \textbf{\# SPGs}\\
\cline{1-3}
\endhead
\cline{1-3}
\endfoot
\hline
\endlastfoot
$\mathcal{C}_1$ & $\varnothing$ & $\hat{R}_{1}:$ \ $\bigl(-R_{[100]}\!\left(\frac{\pi}{2}\right)\Vert T\bigr)$ & 1 \\ \hline
$\mathcal{C}_i$ & $\mathcal{I}$ & $\hat{R}_{1}:$ \ $\bigl(E\Vert \mathcal{I}\bigr)$,\allowbreak\ $\bigl(-R_{[100]}\!\left(\frac{\pi}{2}\right)\Vert T\bigr)$ & 2 \\ \cline{1-3}
$\mathcal{C}_i$ & $\mathcal{I}$ & $\hat{R}_{2}:$ \ $\bigl(R_{[100]}\!\left(\frac{\pi}{2}\right)\Vert \mathcal{I}\bigr)$,\allowbreak\ $\bigl(-R_{[100]}\!\left(\frac{\pi}{2}\right)\Vert T\bigr)$ &  \\ \hline
$\mathcal{C}_2$ & $C_{2z}$ & $\hat{R}_{1}:$ \ $\bigl(E\Vert C_{2z}\bigr)$,\allowbreak\ $\bigl(-R_{[100]}\!\left(\frac{\pi}{2}\right)\Vert T\bigr)$ & 2 \\ \cline{1-3}
$\mathcal{C}_2$ & $C_{2z}$ & $\hat{R}_{2}:$ \ $\bigl(R_{[100]}\!\left(\frac{\pi}{2}\right)\Vert C_{2z}\bigr)$,\allowbreak\ $\bigl(-R_{[100]}\!\left(\frac{\pi}{2}\right)\Vert T\bigr)$ &  \\ \hline
$\mathcal{C}_s$ & $\mathcal{M}_x$ & $\hat{R}_{1}:$ \ $\bigl(E\Vert \mathcal{M}_x\bigr)$,\allowbreak\ $\bigl(-R_{[100]}\!\left(\frac{\pi}{2}\right)\Vert T\bigr)$ & 2 \\ \cline{1-3}
$\mathcal{C}_s$ & $\mathcal{M}_x$ & $\hat{R}_{2}:$ \ $\bigl(R_{[100]}\!\left(\frac{\pi}{2}\right)\Vert \mathcal{M}_x\bigr)$,\allowbreak\ $\bigl(-R_{[100]}\!\left(\frac{\pi}{2}\right)\Vert T\bigr)$ &  \\ \hline
$\mathcal{C}_{2h}$ & $C_{2z},\ \mathcal{I}$ & $\hat{R}_{1}:$ \ $\bigl(E\Vert C_{2z}\bigr)$,\allowbreak\ $\bigl(E\Vert \mathcal{I}\bigr)$,\allowbreak\ $\bigl(-R_{[100]}\!\left(\frac{\pi}{2}\right)\Vert T\bigr)$ &  \\
\hline
\end{longtable}
}

\newpage

\Needspace{8\baselineskip}
\subsection{$S_0=\mathcal{D}_3$}
\subsubsection{1. $S_0=\mathcal{D}_3$: SPGs with exact time-reversal symmetry }

{\scriptsize
\setlength{\LTpre}{0pt}
\setlength{\LTpost}{0pt}
\begin{longtable}{|L{0.10\PGTwidth}|L{0.18\PGTwidth}|L{0.64\PGTwidth}|C{0.08\PGTwidth}|}
\hline
\rowcolor{gray!12}\textbf{PG} & \textbf{PG generators} & \textbf{Representatives (on generators)} & \textbf{\# SPGs}\\
\cline{1-3}
\endfirsthead
\hline
\rowcolor{gray!12}\textbf{PG} & \textbf{PG generators} & \textbf{Representatives (on generators)} & \textbf{\# SPGs}\\
\cline{1-3}
\endhead
\cline{1-3}
\endfoot
\hline
\endlastfoot
$\mathcal{C}_1$ & $\varnothing$ & $\hat{R}_{1}:$ \ (trivial),\allowbreak\ $\bigl(-E\Vert T\bigr)$ & 1 \\ \hline
$\mathcal{C}_i$ & $\mathcal{I}$ & $\hat{R}_{1}:$ \ $\bigl(E\Vert \mathcal{I}\bigr)$,\allowbreak\ $\bigl(-E\Vert T\bigr)$ & 2 \\ \cline{1-3}
$\mathcal{C}_i$ & $\mathcal{I}$ & $\hat{R}_{2}:$ \ $\bigl(R_{[001]}\!\left(\frac{\pi}{3}\right)\Vert \mathcal{I}\bigr)$,\allowbreak\ $\bigl(-E\Vert T\bigr)$ &  \\ \hline
$\mathcal{C}_2$ & $C_{2z}$ & $\hat{R}_{1}:$ \ $\bigl(E\Vert C_{2z}\bigr)$,\allowbreak\ $\bigl(-E\Vert T\bigr)$ & 2 \\ \cline{1-3}
$\mathcal{C}_2$ & $C_{2z}$ & $\hat{R}_{2}:$ \ $\bigl(R_{[001]}\!\left(\frac{\pi}{3}\right)\Vert C_{2z}\bigr)$,\allowbreak\ $\bigl(-E\Vert T\bigr)$ &  \\ \hline
$\mathcal{C}_s$ & $\mathcal{M}_x$ & $\hat{R}_{1}:$ \ $\bigl(E\Vert \mathcal{M}_x\bigr)$,\allowbreak\ $\bigl(-E\Vert T\bigr)$ & 2 \\ \cline{1-3}
$\mathcal{C}_s$ & $\mathcal{M}_x$ & $\hat{R}_{2}:$ \ $\bigl(R_{[001]}\!\left(\frac{\pi}{3}\right)\Vert \mathcal{M}_x\bigr)$,\allowbreak\ $\bigl(-E\Vert T\bigr)$ &  \\ \hline
$\mathcal{C}_{2h}$ & $C_{2z},\ \mathcal{I}$ & $\hat{R}_{1}:$ \ $\bigl(E\Vert C_{2z}\bigr)$,\allowbreak\ $\bigl(E\Vert \mathcal{I}\bigr)$,\allowbreak\ $\bigl(-E\Vert T\bigr)$ &  \\
\hline
\end{longtable}
}

\newpage
\subsubsection{2. $S_0=\mathcal{D}_3$: SPGs with effective time-reversal symmetry }

{\scriptsize
\setlength{\LTpre}{0pt}
\setlength{\LTpost}{0pt}
\begin{longtable}{|L{0.10\PGTwidth}|L{0.18\PGTwidth}|L{0.64\PGTwidth}|C{0.08\PGTwidth}|}
\hline
\rowcolor{gray!12}\textbf{PG} & \textbf{PG generators} & \textbf{Representatives (on generators)} & \textbf{\# SPGs}\\
\cline{1-3}
\endfirsthead
\hline
\rowcolor{gray!12}\textbf{PG} & \textbf{PG generators} & \textbf{Representatives (on generators)} & \textbf{\# SPGs}\\
\cline{1-3}
\endhead
\cline{1-3}
\endfoot
\hline
\endlastfoot
$\mathcal{C}_1$ & $\varnothing$ & $\hat{R}_{1}:$ \ $\bigl(-R_{[001]}\!\left(\frac{\pi}{3}\right)\Vert T\bigr)$ & 1 \\ \hline
$\mathcal{C}_i$ & $\mathcal{I}$ & $\hat{R}_{1}:$ \ $\bigl(E\Vert \mathcal{I}\bigr)$,\allowbreak\ $\bigl(-R_{[001]}\!\left(\frac{\pi}{3}\right)\Vert T\bigr)$ & 2 \\ \cline{1-3}
$\mathcal{C}_i$ & $\mathcal{I}$ & $\hat{R}_{2}:$ \ $\bigl(R_{[001]}\!\left(\frac{\pi}{3}\right)\Vert \mathcal{I}\bigr)$,\allowbreak\ $\bigl(-R_{[001]}\!\left(\frac{\pi}{3}\right)\Vert T\bigr)$ &  \\ \hline
$\mathcal{C}_2$ & $C_{2z}$ & $\hat{R}_{1}:$ \ $\bigl(E\Vert C_{2z}\bigr)$,\allowbreak\ $\bigl(-R_{[001]}\!\left(\frac{\pi}{3}\right)\Vert T\bigr)$ & 2 \\ \cline{1-3}
$\mathcal{C}_2$ & $C_{2z}$ & $\hat{R}_{2}:$ \ $\bigl(R_{[001]}\!\left(\frac{\pi}{3}\right)\Vert C_{2z}\bigr)$,\allowbreak\ $\bigl(-R_{[001]}\!\left(\frac{\pi}{3}\right)\Vert T\bigr)$ &  \\ \hline
$\mathcal{C}_s$ & $\mathcal{M}_x$ & $\hat{R}_{1}:$ \ $\bigl(E\Vert \mathcal{M}_x\bigr)$,\allowbreak\ $\bigl(-R_{[001]}\!\left(\frac{\pi}{3}\right)\Vert T\bigr)$ & 2 \\ \cline{1-3}
$\mathcal{C}_s$ & $\mathcal{M}_x$ & $\hat{R}_{2}:$ \ $\bigl(R_{[001]}\!\left(\frac{\pi}{3}\right)\Vert \mathcal{M}_x\bigr)$,\allowbreak\ $\bigl(-R_{[001]}\!\left(\frac{\pi}{3}\right)\Vert T\bigr)$ &  \\ \hline
$\mathcal{C}_{2h}$ & $C_{2z},\ \mathcal{I}$ & $\hat{R}_{1}:$ \ $\bigl(E\Vert C_{2z}\bigr)$,\allowbreak\ $\bigl(E\Vert \mathcal{I}\bigr)$,\allowbreak\ $\bigl(-R_{[001]}\!\left(\frac{\pi}{3}\right)\Vert T\bigr)$ &  \\
\hline
\end{longtable}
}

\newpage

\Needspace{8\baselineskip}
\subsection{$S_0=\mathcal{D}_4$}
\subsubsection{1. $S_0=\mathcal{D}_4$: SPGs with exact time-reversal symmetry }

{\scriptsize
\setlength{\LTpre}{0pt}
\setlength{\LTpost}{0pt}
\begin{longtable}{|L{0.10\PGTwidth}|L{0.18\PGTwidth}|L{0.64\PGTwidth}|C{0.08\PGTwidth}|}
\hline
\rowcolor{gray!12}\textbf{PG} & \textbf{PG generators} & \textbf{Representatives (on generators)} & \textbf{\# SPGs}\\
\cline{1-3}
\endfirsthead
\hline
\rowcolor{gray!12}\textbf{PG} & \textbf{PG generators} & \textbf{Representatives (on generators)} & \textbf{\# SPGs}\\
\cline{1-3}
\endhead
\cline{1-3}
\endfoot
\hline
\endlastfoot
$\mathcal{C}_1$ & $\varnothing$ & $\hat{R}_{1}:$ \ (trivial),\allowbreak\ $\bigl(-E\Vert T\bigr)$ & 1 \\ \hline
$\mathcal{C}_i$ & $\mathcal{I}$ & $\hat{R}_{1}:$ \ $\bigl(E\Vert \mathcal{I}\bigr)$,\allowbreak\ $\bigl(-E\Vert T\bigr)$ & 2 \\ \cline{1-3}
$\mathcal{C}_i$ & $\mathcal{I}$ & $\hat{R}_{2}:$ \ $\bigl(R_{[001]}\!\left(\frac{\pi}{4}\right)\Vert \mathcal{I}\bigr)$,\allowbreak\ $\bigl(-E\Vert T\bigr)$ &  \\ \hline
$\mathcal{C}_2$ & $C_{2z}$ & $\hat{R}_{1}:$ \ $\bigl(E\Vert C_{2z}\bigr)$,\allowbreak\ $\bigl(-E\Vert T\bigr)$ & 2 \\ \cline{1-3}
$\mathcal{C}_2$ & $C_{2z}$ & $\hat{R}_{2}:$ \ $\bigl(R_{[001]}\!\left(\frac{\pi}{4}\right)\Vert C_{2z}\bigr)$,\allowbreak\ $\bigl(-E\Vert T\bigr)$ &  \\ \hline
$\mathcal{C}_s$ & $\mathcal{M}_x$ & $\hat{R}_{1}:$ \ $\bigl(E\Vert \mathcal{M}_x\bigr)$,\allowbreak\ $\bigl(-E\Vert T\bigr)$ & 2 \\ \cline{1-3}
$\mathcal{C}_s$ & $\mathcal{M}_x$ & $\hat{R}_{2}:$ \ $\bigl(R_{[001]}\!\left(\frac{\pi}{4}\right)\Vert \mathcal{M}_x\bigr)$,\allowbreak\ $\bigl(-E\Vert T\bigr)$ &  \\ \hline
$\mathcal{C}_{2h}$ & $C_{2z},\ \mathcal{I}$ & $\hat{R}_{1}:$ \ $\bigl(E\Vert C_{2z}\bigr)$,\allowbreak\ $\bigl(E\Vert \mathcal{I}\bigr)$,\allowbreak\ $\bigl(-E\Vert T\bigr)$ &  \\
\hline
\end{longtable}
}

\newpage
\subsubsection{2. $S_0=\mathcal{D}_4$: SPGs with effective time-reversal symmetry }

{\scriptsize
\setlength{\LTpre}{0pt}
\setlength{\LTpost}{0pt}
\begin{longtable}{|L{0.10\PGTwidth}|L{0.18\PGTwidth}|L{0.64\PGTwidth}|C{0.08\PGTwidth}|}
\hline
\rowcolor{gray!12}\textbf{PG} & \textbf{PG generators} & \textbf{Representatives (on generators)} & \textbf{\# SPGs}\\
\cline{1-3}
\endfirsthead
\hline
\rowcolor{gray!12}\textbf{PG} & \textbf{PG generators} & \textbf{Representatives (on generators)} & \textbf{\# SPGs}\\
\cline{1-3}
\endhead
\cline{1-3}
\endfoot
\hline
\endlastfoot
$\mathcal{C}_1$ & $\varnothing$ & $\hat{R}_{1}:$ \ $\bigl(-R_{[001]}\!\left(\frac{\pi}{4}\right)\Vert T\bigr)$ & 1 \\ \hline
$\mathcal{C}_i$ & $\mathcal{I}$ & $\hat{R}_{1}:$ \ $\bigl(E\Vert \mathcal{I}\bigr)$,\allowbreak\ $\bigl(-R_{[001]}\!\left(\frac{\pi}{4}\right)\Vert T\bigr)$ & 2 \\ \cline{1-3}
$\mathcal{C}_i$ & $\mathcal{I}$ & $\hat{R}_{2}:$ \ $\bigl(R_{[001]}\!\left(\frac{\pi}{4}\right)\Vert \mathcal{I}\bigr)$,\allowbreak\ $\bigl(-R_{[001]}\!\left(\frac{\pi}{4}\right)\Vert T\bigr)$ &  \\ \hline
$\mathcal{C}_2$ & $C_{2z}$ & $\hat{R}_{1}:$ \ $\bigl(E\Vert C_{2z}\bigr)$,\allowbreak\ $\bigl(-R_{[001]}\!\left(\frac{\pi}{4}\right)\Vert T\bigr)$ & 2 \\ \cline{1-3}
$\mathcal{C}_2$ & $C_{2z}$ & $\hat{R}_{2}:$ \ $\bigl(R_{[001]}\!\left(\frac{\pi}{4}\right)\Vert C_{2z}\bigr)$,\allowbreak\ $\bigl(-R_{[001]}\!\left(\frac{\pi}{4}\right)\Vert T\bigr)$ &  \\ \hline
$\mathcal{C}_s$ & $\mathcal{M}_x$ & $\hat{R}_{1}:$ \ $\bigl(E\Vert \mathcal{M}_x\bigr)$,\allowbreak\ $\bigl(-R_{[001]}\!\left(\frac{\pi}{4}\right)\Vert T\bigr)$ & 2 \\ \cline{1-3}
$\mathcal{C}_s$ & $\mathcal{M}_x$ & $\hat{R}_{2}:$ \ $\bigl(R_{[001]}\!\left(\frac{\pi}{4}\right)\Vert \mathcal{M}_x\bigr)$,\allowbreak\ $\bigl(-R_{[001]}\!\left(\frac{\pi}{4}\right)\Vert T\bigr)$ &  \\ \hline
$\mathcal{C}_{2h}$ & $C_{2z},\ \mathcal{I}$ & $\hat{R}_{1}:$ \ $\bigl(E\Vert C_{2z}\bigr)$,\allowbreak\ $\bigl(E\Vert \mathcal{I}\bigr)$,\allowbreak\ $\bigl(-R_{[001]}\!\left(\frac{\pi}{4}\right)\Vert T\bigr)$ &  \\
\hline
\end{longtable}
}

\newpage
\Needspace{8\baselineskip}
\subsection{$S_0=\mathcal{D}_6$}
\subsubsection{1. $S_0=\mathcal{D}_6$: SPGs with exact time-reversal symmetry }
{\scriptsize
\setlength{\LTpre}{0pt}
\setlength{\LTpost}{0pt}
\begin{longtable}{|L{0.10\PGTwidth}|L{0.18\PGTwidth}|L{0.64\PGTwidth}|C{0.08\PGTwidth}|}
\hline
\rowcolor{gray!12}\textbf{PG} & \textbf{PG generators} & \textbf{Representatives (on generators)} & \textbf{\# SPGs}\\
\cline{1-3}
\endfirsthead
\hline
\rowcolor{gray!12}\textbf{PG} & \textbf{PG generators} & \textbf{Representatives (on generators)} & \textbf{\# SPGs}\\
\cline{1-3}
\endhead
\cline{1-3}
\endfoot
\hline
\endlastfoot
$\mathcal{C}_1$ & $\varnothing$ & $\hat{R}_{1}:$ \ (trivial),\allowbreak\ $\bigl(-E\Vert T\bigr)$ & 1 \\ \hline
$\mathcal{C}_i$ & $\mathcal{I}$ & $\hat{R}_{1}:$ \ $\bigl(E\Vert \mathcal{I}\bigr)$,\allowbreak\ $\bigl(-E\Vert T\bigr)$ & 2 \\ \cline{1-3}
$\mathcal{C}_i$ & $\mathcal{I}$ & $\hat{R}_{2}:$ \ $\bigl(R_{[001]}\!\left(\frac{\pi}{6}\right)\Vert \mathcal{I}\bigr)$,\allowbreak\ $\bigl(-E\Vert T\bigr)$ &  \\ \hline
$\mathcal{C}_2$ & $C_{2z}$ & $\hat{R}_{1}:$ \ $\bigl(E\Vert C_{2z}\bigr)$,\allowbreak\ $\bigl(-E\Vert T\bigr)$ & 2 \\ \cline{1-3}
$\mathcal{C}_2$ & $C_{2z}$ & $\hat{R}_{2}:$ \ $\bigl(R_{[001]}\!\left(\frac{\pi}{6}\right)\Vert C_{2z}\bigr)$,\allowbreak\ $\bigl(-E\Vert T\bigr)$ &  \\ \hline
$\mathcal{C}_s$ & $\mathcal{M}_x$ & $\hat{R}_{1}:$ \ $\bigl(E\Vert \mathcal{M}_x\bigr)$,\allowbreak\ $\bigl(-E\Vert T\bigr)$ & 2 \\ \cline{1-3}
$\mathcal{C}_s$ & $\mathcal{M}_x$ & $\hat{R}_{2}:$ \ $\bigl(R_{[001]}\!\left(\frac{\pi}{6}\right)\Vert \mathcal{M}_x\bigr)$,\allowbreak\ $\bigl(-E\Vert T\bigr)$ &  \\ \hline
$\mathcal{C}_{2h}$ & $C_{2z},\ \mathcal{I}$ & $\hat{R}_{1}:$ \ $\bigl(E\Vert C_{2z}\bigr)$,\allowbreak\ $\bigl(E\Vert \mathcal{I}\bigr)$,\allowbreak\ $\bigl(-E\Vert T\bigr)$ &  \\
\hline
\end{longtable}
}

\newpage
\subsubsection{2. $S_0=\mathcal{D}_6$: SPGs with effective time-reversal symmetry }
{\scriptsize
\setlength{\LTpre}{0pt}
\setlength{\LTpost}{0pt}
\begin{longtable}{|L{0.10\PGTwidth}|L{0.18\PGTwidth}|L{0.64\PGTwidth}|C{0.08\PGTwidth}|}
\hline
\rowcolor{gray!12}\textbf{PG} & \textbf{PG generators} & \textbf{Representatives (on generators)} & \textbf{\# SPGs}\\
\cline{1-3}
\endfirsthead
\hline
\rowcolor{gray!12}\textbf{PG} & \textbf{PG generators} & \textbf{Representatives (on generators)} & \textbf{\# SPGs}\\
\cline{1-3}
\endhead
\cline{1-3}
\endfoot
\hline
\endlastfoot
$\mathcal{C}_1$ & $\varnothing$ & $\hat{R}_{1}:$ \ $\bigl(-R_{[001]}\!\left(\frac{\pi}{6}\right)\Vert T\bigr)$ & 1 \\ \hline
$\mathcal{C}_i$ & $\mathcal{I}$ & $\hat{R}_{1}:$ \ $\bigl(E\Vert \mathcal{I}\bigr)$,\allowbreak\ $\bigl(-R_{[001]}\!\left(\frac{\pi}{6}\right)\Vert T\bigr)$ & 2 \\ \cline{1-3}
$\mathcal{C}_i$ & $\mathcal{I}$ & $\hat{R}_{2}:$ \ $\bigl(R_{[001]}\!\left(\frac{\pi}{6}\right)\Vert \mathcal{I}\bigr)$,\allowbreak\ $\bigl(-R_{[001]}\!\left(\frac{\pi}{6}\right)\Vert T\bigr)$ &  \\ \hline
$\mathcal{C}_2$ & $C_{2z}$ & $\hat{R}_{1}:$ \ $\bigl(E\Vert C_{2z}\bigr)$,\allowbreak\ $\bigl(-R_{[001]}\!\left(\frac{\pi}{6}\right)\Vert T\bigr)$ & 2 \\ \cline{1-3}
$\mathcal{C}_2$ & $C_{2z}$ & $\hat{R}_{2}:$ \ $\bigl(R_{[001]}\!\left(\frac{\pi}{6}\right)\Vert C_{2z}\bigr)$,\allowbreak\ $\bigl(-R_{[001]}\!\left(\frac{\pi}{6}\right)\Vert T\bigr)$ &  \\ \hline
$\mathcal{C}_s$ & $\mathcal{M}_x$ & $\hat{R}_{1}:$ \ $\bigl(E\Vert \mathcal{M}_x\bigr)$,\allowbreak\ $\bigl(-R_{[001]}\!\left(\frac{\pi}{6}\right)\Vert T\bigr)$ & 2 \\ \cline{1-3}
$\mathcal{C}_s$ & $\mathcal{M}_x$ & $\hat{R}_{2}:$ \ $\bigl(R_{[001]}\!\left(\frac{\pi}{6}\right)\Vert \mathcal{M}_x\bigr)$,\allowbreak\ $\bigl(-R_{[001]}\!\left(\frac{\pi}{6}\right)\Vert T\bigr)$ &  \\ \hline
$\mathcal{C}_{2h}$ & $C_{2z},\ \mathcal{I}$ & $\hat{R}_{1}:$ \ $\bigl(E\Vert C_{2z}\bigr)$,\allowbreak\ $\bigl(E\Vert \mathcal{I}\bigr)$,\allowbreak\ $\bigl(-R_{[001]}\!\left(\frac{\pi}{6}\right)\Vert T\bigr)$ &  \\
\hline
\end{longtable}
}

\newpage
\Needspace{8\baselineskip}
\subsection{$S_0=\mathcal{T}$}
\subsubsection{1. $S_0=\mathcal{T}$: SPGs with exact time-reversal symmetry }
{\scriptsize
\setlength{\LTpre}{0pt}
\setlength{\LTpost}{0pt}
\begin{longtable}{|L{0.10\PGTwidth}|L{0.18\PGTwidth}|L{0.64\PGTwidth}|C{0.08\PGTwidth}|}
\hline
\rowcolor{gray!12}\textbf{PG} & \textbf{PG generators} & \textbf{Representatives (on generators)} & \textbf{\# SPGs}\\
\cline{1-3}
\endfirsthead
\hline
\rowcolor{gray!12}\textbf{PG} & \textbf{PG generators} & \textbf{Representatives (on generators)} & \textbf{\# SPGs}\\
\cline{1-3}
\endhead
\cline{1-3}
\endfoot
\hline
\endlastfoot
$\mathcal{C}_1$ & $\varnothing$ & $\hat{R}_{1}:$ \ (trivial),\allowbreak\ $\bigl(-E\Vert T\bigr)$ & 1 \\ \hline
$\mathcal{C}_i$ & $\mathcal{I}$ & $\hat{R}_{1}:$ \ $\bigl(E\Vert \mathcal{I}\bigr)$,\allowbreak\ $\bigl(-E\Vert T\bigr)$ & 2 \\ \cline{1-3}
$\mathcal{C}_i$ & $\mathcal{I}$ & $\hat{R}_{2}:$ \ $\bigl(R_{[001]}\!\left(\frac{\pi}{2}\right)\Vert \mathcal{I}\bigr)$,\allowbreak\ $\bigl(-E\Vert T\bigr)$ &  \\ \hline
$\mathcal{C}_2$ & $C_{2z}$ & $\hat{R}_{1}:$ \ $\bigl(E\Vert C_{2z}\bigr)$,\allowbreak\ $\bigl(-E\Vert T\bigr)$ & 2 \\ \cline{1-3}
$\mathcal{C}_2$ & $C_{2z}$ & $\hat{R}_{2}:$ \ $\bigl(R_{[001]}\!\left(\frac{\pi}{2}\right)\Vert C_{2z}\bigr)$,\allowbreak\ $\bigl(-E\Vert T\bigr)$ &  \\ \hline
$\mathcal{C}_s$ & $\mathcal{M}_x$ & $\hat{R}_{1}:$ \ $\bigl(E\Vert \mathcal{M}_x\bigr)$,\allowbreak\ $\bigl(-E\Vert T\bigr)$ & 2 \\ \cline{1-3}
$\mathcal{C}_s$ & $\mathcal{M}_x$ & $\hat{R}_{2}:$ \ $\bigl(R_{[001]}\!\left(\frac{\pi}{2}\right)\Vert \mathcal{M}_x\bigr)$,\allowbreak\ $\bigl(-E\Vert T\bigr)$ &  \\ \hline
$\mathcal{C}_{2h}$ & $C_{2z},\ \mathcal{I}$ & $\hat{R}_{1}:$ \ $\bigl(E\Vert C_{2z}\bigr)$,\allowbreak\ $\bigl(E\Vert \mathcal{I}\bigr)$,\allowbreak\ $\bigl(-E\Vert T\bigr)$ &  \\
\hline
\end{longtable}
}

\newpage
\subsubsection{2. $S_0=\mathcal{T}$: SPGs with effective time-reversal symmetry }
{\scriptsize
\setlength{\LTpre}{0pt}
\setlength{\LTpost}{0pt}
\begin{longtable}{|L{0.10\PGTwidth}|L{0.18\PGTwidth}|L{0.64\PGTwidth}|C{0.08\PGTwidth}|}
\hline
\rowcolor{gray!12}\textbf{PG} & \textbf{PG generators} & \textbf{Representatives (on generators)} & \textbf{\# SPGs}\\
\cline{1-3}
\endfirsthead
\hline
\rowcolor{gray!12}\textbf{PG} & \textbf{PG generators} & \textbf{Representatives (on generators)} & \textbf{\# SPGs}\\
\cline{1-3}
\endhead
\cline{1-3}
\endfoot
\hline
\endlastfoot
$\mathcal{C}_1$ & $\varnothing$ & $\hat{R}_{1}:$ \ $\bigl(-R_{[001]}\!\left(\frac{\pi}{2}\right)\Vert T\bigr)$ & 1 \\ \hline
$\mathcal{C}_i$ & $\mathcal{I}$ & $\hat{R}_{1}:$ \ $\bigl(E\Vert \mathcal{I}\bigr)$,\allowbreak\ $\bigl(-R_{[001]}\!\left(\frac{\pi}{2}\right)\Vert T\bigr)$ & 2 \\ \cline{1-3}
$\mathcal{C}_i$ & $\mathcal{I}$ & $\hat{R}_{2}:$ \ $\bigl(R_{[001]}\!\left(\frac{\pi}{2}\right)\Vert \mathcal{I}\bigr)$,\allowbreak\ $\bigl(-R_{[001]}\!\left(\frac{\pi}{2}\right)\Vert T\bigr)$ &  \\ \hline
$\mathcal{C}_2$ & $C_{2z}$ & $\hat{R}_{1}:$ \ $\bigl(E\Vert C_{2z}\bigr)$,\allowbreak\ $\bigl(-R_{[001]}\!\left(\frac{\pi}{2}\right)\Vert T\bigr)$ & 2 \\ \cline{1-3}
$\mathcal{C}_2$ & $C_{2z}$ & $\hat{R}_{2}:$ \ $\bigl(R_{[001]}\!\left(\frac{\pi}{2}\right)\Vert C_{2z}\bigr)$,\allowbreak\ $\bigl(-R_{[001]}\!\left(\frac{\pi}{2}\right)\Vert T\bigr)$ &  \\ \hline
$\mathcal{C}_s$ & $\mathcal{M}_x$ & $\hat{R}_{1}:$ \ $\bigl(E\Vert \mathcal{M}_x\bigr)$,\allowbreak\ $\bigl(-R_{[001]}\!\left(\frac{\pi}{2}\right)\Vert T\bigr)$ & 2 \\ \cline{1-3}
$\mathcal{C}_s$ & $\mathcal{M}_x$ & $\hat{R}_{2}:$ \ $\bigl(R_{[001]}\!\left(\frac{\pi}{2}\right)\Vert \mathcal{M}_x\bigr)$,\allowbreak\ $\bigl(-R_{[001]}\!\left(\frac{\pi}{2}\right)\Vert T\bigr)$ &  \\ \hline
$\mathcal{C}_{2h}$ & $C_{2z},\ \mathcal{I}$ & $\hat{R}_{1}:$ \ $\bigl(E\Vert C_{2z}\bigr)$,\allowbreak\ $\bigl(E\Vert \mathcal{I}\bigr)$,\allowbreak\ $\bigl(-R_{[001]}\!\left(\frac{\pi}{2}\right)\Vert T\bigr)$ &  \\
\hline
\end{longtable}
}

\newpage

\Needspace{8\baselineskip}
\subsection{$S_0=\mathcal{O}$}
\subsubsection{1. $S_0=\mathcal{O}$: SPGs with exact time-reversal symmetry }
{\scriptsize
\setlength{\LTpre}{0pt}
\setlength{\LTpost}{0pt}
\begin{longtable}{|L{0.10\PGTwidth}|L{0.18\PGTwidth}|L{0.64\PGTwidth}|C{0.08\PGTwidth}|}
\hline
\rowcolor{gray!12}\textbf{PG} & \textbf{PG generators} & \textbf{Representatives (on generators)} & \textbf{\# SPGs}\\
\cline{1-3}
\endfirsthead
\hline
\rowcolor{gray!12}\textbf{PG} & \textbf{PG generators} & \textbf{Representatives (on generators)} & \textbf{\# SPGs}\\
\cline{1-3}
\endhead
\cline{1-3}
\endfoot
\hline
\endlastfoot
$\mathcal{C}_1$ & $\varnothing$ & $\hat{R}_{1}:$ \ (trivial),\allowbreak\ $\bigl(-E\Vert T\bigr)$ & 1 \\ \hline
$\mathcal{C}_i$ & $\mathcal{I}$ & $\hat{R}_{1}:$ \ $\bigl(E\Vert \mathcal{I}\bigr)$,\allowbreak\ $\bigl(-E\Vert T\bigr)$ & 1 \\ \hline
$\mathcal{C}_2$ & $C_{2z}$ & $\hat{R}_{1}:$ \ $\bigl(E\Vert C_{2z}\bigr)$,\allowbreak\ $\bigl(-E\Vert T\bigr)$ & 1 \\ \hline
$\mathcal{C}_s$ & $\mathcal{M}_x$ & $\hat{R}_{1}:$ \ $\bigl(E\Vert \mathcal{M}_x\bigr)$,\allowbreak\ $\bigl(-E\Vert T\bigr)$ & 1 \\ \hline
$\mathcal{C}_{2h}$ & $C_{2z},\ \mathcal{I}$ & $\hat{R}_{1}:$ \ $\bigl(E\Vert C_{2z}\bigr)$,\allowbreak\ $\bigl(E\Vert \mathcal{I}\bigr)$,\allowbreak\ $\bigl(-E\Vert T\bigr)$ & 1 \\ \hline
$\mathcal{D}_2$ & $C_{2x},\ C_{2y}$ & $\hat{R}_{1}:$ \ $\bigl(E\Vert C_{2x}\bigr)$,\allowbreak\ $\bigl(E\Vert C_{2y}\bigr)$,\allowbreak\ $\bigl(-E\Vert T\bigr)$ & 1 \\ \hline
$\mathcal{C}_{2v}$ & $C_{2z},\ \mathcal{M}_x$ & $\hat{R}_{1}:$ \ $\bigl(E\Vert C_{2z}\bigr)$,\allowbreak\ $\bigl(E\Vert \mathcal{M}_x\bigr)$,\allowbreak\ $\bigl(-E\Vert T\bigr)$ & 1 \\ \hline
$\mathcal{D}_{2h}$ & $\mathcal{M}_x,\ \mathcal{M}_y,\ \mathcal{M}_z$ & $\hat{R}_{1}:$ \ $\bigl(E\Vert \mathcal{M}_x\bigr)$,\allowbreak\ $\bigl(E\Vert \mathcal{M}_y\bigr)$,\allowbreak\ $\bigl(E\Vert \mathcal{M}_z\bigr)$,\allowbreak\ $\bigl(-E\Vert T\bigr)$ & 1 \\
\hline
\end{longtable}
}

\IfFileExists{tSSGs_references.bib}{%
  \bibliographystyle{apsrev}%
  \bibliography{tSSGs_references}%
}{}